\documentclass{aa}  

\usepackage{graphicx}
\usepackage{txfonts}
\usepackage{xcolor}
\usepackage{gensymb}
\usepackage{figsize}
\usepackage{url}

\usepackage{hyperref} 
\hypersetup{breaklinks=true,colorlinks=true,linkcolor=blue,citecolor=blue,urlcolor=blue}
\usepackage{natbib,twoopt}
\bibpunct{(}{)}{;}{a}{}{,}

\makeatletter

\newcommandtwoopt{\citeads}[3][][]{\href{http://adsabs.harvard.edu/abs/#3}%
{\def\hyper@linkstart##1##2{}%
\let\hyper@linkend\@empty\citealp[#1][#2]{#3}}}
\newcommandtwoopt{\citepads}[3][][]{\href{http://adsabs.harvard.edu/abs/#3}%
{\def\hyper@linkstart##1##2{}%
\let\hyper@linkend\@empty\citep[#1][#2]{#3}}}
\newcommandtwoopt{\citetads}[3][][]{\href{http://adsabs.harvard.edu/abs/#3}%
{\def\hyper@linkstart##1##2{}%
\let\hyper@linkend\@empty\citet[#1][#2]{#3}}}
\newcommandtwoopt{\citeyearads}[3][][]%
{\href{http://adsabs.harvard.edu/abs/#3}
{\def\hyper@linkstart##1##2{}%
\let\hyper@linkend\@empty\citeyear[#1][#2]{#3}}}
\newcommandtwoopt{\citealtads}[3][][]%
{\href{http://adsabs.harvard.edu/abs/#3}
{\def\hyper@linkstart##1##2{}%
\let\hyper@linkend\@empty\citealt[#1][#2]{#3}}}
\makeatother

\newcommand{\kms}{km\,s$^{-1}$}

\def\deltaV#1{$\Delta v_\star$}
\def\deltaS#1{$\Delta \sigma_\star$}

\begin{document}

   \title{The kinematics of young and old stellar populations in nuclear rings of MUSE TIMER galaxies}


   \author{D. Rosado-Belza
          \inst{1,}
          \inst{2},
          J. Falc\'{o}n-Barroso
          \inst{1,}
          \inst{2},
          J. H. Knapen
          \inst{1,}
          \inst{2},
          A. Bittner
          \inst{3,}
          \inst{4},
          D. A. Gadotti
          \inst{3},
          J. Neumann
          \inst{5},
          A. de Lorenzo-C\'{a}ceres
          \inst{1,}
          \inst{2},
          J. M\'{e}ndez-Abreu
          \inst{1,}
          \inst{2,}
          \inst{6,}
          \inst{7},
          M. Querejeta
          \inst{3,}
          \inst{8},
          I. Mart\'{i}n-Navarro
          \inst{1,}
          \inst{2},
          P. S\'{a}nchez-Bl\'{a}zquez
          \inst{9,}
          \inst{10},
          P. R. T. Coelho
          \inst{11},
          M. Martig
          \inst{12},
          G. van de Ven
          \inst{13},
          \and
          T. Kim
          \inst{14}
          }
          
   \institute{Instituto de Astrof\'{i}sica de Canarias,
              E-38205, San Crist\'{o}bal de La Laguna, Santa Cruz de Tenerife, Spain,
              \email{drb@iac.es}
         \and
             Departamento de Astrof\'{i}sica, Universidad de La Laguna, E-38205, San Crist\'{o}bal de La Laguna, Santa Cruz de Tenerife, Spain
        \and
            European Southern Observatory, Karl-Schwarzschild-Str. 2, D-85748 Garching bei M\"{u}nchen, Germany
        \and
            Ludwig-Maximilians-Universit\"{a}t, Professor-Huber-Platz 2, 80539 M\"{u}nchen, Germany
        \and
            Institute of Cosmology and Gravitation, University of Portsmouth, Burnaby Road, Portsmouth PO1 3FX, UK
        \and
            Dpto. de F\'{i}sica y del Cosmos, Campus de Fuentenueva, Edificio Mecenas, Universidad de Granada, E18071–Granada, Spain
        \and
            Instituto Carlos I de F\'{i}sica Te\'{o}rica y Computacional, Facultad de Ciencias, E18071–Granada, Spain
        \and
            Observatorio Astron\'{o}mico Nacional (IGN), C/ Alfonso XII 3, 28014 Madrid, Spain
        \and
            Departamento de F\'{i}sica de la Tierra y Astrof\'{i}sica, Universidad Complutense de Madrid, 28040, Madrid, Spain
        \and
            IPARCOS, Facultad de C.C. F\'{i}sicas, Universidad Complutense de Madrid, 28040, Madrid, Spain
        \and
            Instituto de Astronomia, Geof\'{i}sica e Ci\^{e}ncias Atmosf\'{e}ricas, Universidade de S\~{a}o Paulo, S\~{a}o Paulo, SP, Brazil
        \and
            Astrophysics Research Institute, Liverpool John Moores University, 146 Brownlow Hill, L3 5RF Liverpool, UK
        \and
            Department of Astrophysics, University of Vienna
            T\"{u}rkenschanzstra{\ss}e 17, 1180 Vienna, Austria
        \and
            Department of Astronomy and Atmospheric Sciences, Kyungpook National University, Daegu 702-701, Korea
             }
            
   \date{}

 \abstract{Studying the stellar kinematics of galaxies is a key tool in the reconstruction of their evolution. However, the current measurements of the stellar kinematics are complicated by several factors, including dust extinction and the presence of multiple stellar populations.}
{We use integral field spectroscopic data of four galaxies from the Time Inference with MUSE in Extragalactic Rings (TIMER) survey to explore and compare the kinematics measured in different spectral regions that are sensitive to distinct stellar populations.}
{We derive the line-of-sight velocity and velocity dispersion of both a young ($\lesssim 2$\,Gyr) and an old stellar population from the spectral regions around the H$\beta$ line and the Ca\,II Triplet. In addition we obtain colour excess, mean age, and metallicity.}
{We report a correlation of the colour excess with the difference in the kinematic parameters of the H$\beta$ line and the Ca\,II Triplet range, which are dominated by young and old stellar populations, respectively. Young stellar populations, located primarily in nuclear rings, have higher velocity dispersions than old ones. These differences in the rings are typically $\sim10\,{\rm km\,s^{-1}}$ in velocity dispersion, but up to a mean value of $\sim24\,{\rm km\,s^{-1}}$ in the most extreme case. Trends with age exist in the nuclear rings but are less significant than those with dust extinction. We report different degrees of correlation of these trends among the galaxies in the sample, which are related to the size of the Voronoi bins in their rings. No clear trends for the difference of line-of-sight velocity are observed. The absence of these trends can be explained as a consequence of the masking process of the H$\beta$ line during the kinematic extraction, as confirmed by dedicated simulations.}
{Our study demonstrates that kinematic differences caused by different stellar populations can be identified in the central regions of nearby galaxies even from intermediate resolution spectroscopy. This opens the door to future detailed chemo-kinematic studies of galaxies, but also serves as a warning against deriving kinematics from full-spectrum fitting across very wide wavelength ranges when intense star formation is taking place.}

   \keywords{}
    \titlerunning{The kinematics of young and old stellar populations in nuclear rings}
    \authorrunning{D. Rosado-Belza et al.}
   \maketitle
%
\section{Introduction}
\label{sec:intro}

The evolution of galaxies is one of the most important topics in extragalactic astronomy. Galaxies evolve in different ways under the influence of internal or external processes, leading to different morphologies \citepads{2004ARA&A..42..603K}. Internal secular processes are slow, with timescales equivalent to several times the rotation period of the galaxy \citepads[e.g.][]{1979ApJ...227..714K, 1981A&A....96..164C, 1998gaas.book.....B, 2004ARA&A..42..603K, 2013seg..book....1K}, but they are particularly relevant in the low-density regions of the Universe preferred by late-type galaxies.

Massive discs and bars play an important role in secular evolution. Galactic bars are formed in the early stages of a galaxy lifetime \citepads{2012ApJ...758..136S}, shortly after the disc is settled and becomes dynamically cold. The bar behaves like a true engine of evolution of the galaxy, driving the distribution of angular momentum in the disc and causing gas inflow. Disc gas can be funnelled towards the inner regions of a galaxy, where it can lead to new structures or even help fuel AGN activity \citepads[e.g.][]{1985A&A...150..327C, 1989Natur.338...45S,1999ApJ...525..691S,2004ApJ...600..595R,2005ApJ...632..217S,2015MNRAS.449.2421S}. Nuclear rings are stimulated by the gas inflow driven by the bar to the central parts of the galaxy. The gas inflow stagnates near inner Lindblad dynamical resonances where the gas piles up, shocks, and leads to star formation \citepads[e.g.][]{1995ApJ...454..623K, 1995ApJ...449..508P, 1996FCPh...17...95B, 2000A&A...362..465R, 2001AJ....121.3048M, 2006MNRAS.369..529F, 2010MNRAS.402.2462C}. Studying the kinematics of older and younger stellar populations in nuclear rings can provide valuable information on when and how rings and bars were built and thus also on when the disc settled \citepads[][]{2015A&A...584A..90G, 2019MNRAS.482..506G}.\

The evolution of the inner structures of a galaxy can be studied using both theoretical and observational approaches. In the last few decades the theoretical approach has advanced significantly, mainly based on hydrodynamical simulations. These can reproduce the kinematics of a wide variety of galaxies at different redshifts and thus contribute to the development of a general understanding of galaxy evolution. Models like EAGLE \citepads{2015MNRAS.446..521S} and IllustrisTNG \citepads{2018MNRAS.473.4077P} have become very popular and successfully replicate the results of key observations.\looseness-2
 
On the observational side, technical advances have resulted in tremendous progress. One of these is the introduction of integral-field spectroscopy (IFS), from which spatially resolved maps of gas and stellar kinematics in galaxies can be derived. Combining IFS data with stellar population models has allowed the study of the distribution of kinematics, age, metallicity, extinction, and colour of the stellar populations in different parts of galaxies at unprecedented resolution. From the first implementations with instruments such as OASIS or SAURON \citepads[e.g.][]{2001MNRAS.326...23B}, IFS has evolved into revolutionary instruments like VIMOS \citepads[][]{2003SPIE.4841.1670L}, SINFONI \citepads[][]{2003SPIE.4841.1548E, 2004Msngr.117...17B}, PMAS/PPaK \citepads[][]{2005PASP..117..620R, 2004AN....325..151V, 2006PASP..118..129K}, MUSE \citepads[][]{2010SPIE.7735E..08B}, or MEGARA \citepads[][]{2014SPIE.9147E..0OG, 2018SPIE10702E..17G, 2018SPIE10702E..16C}, and into surveys like ATLAS3D \citepads[][]{2011MNRAS.413..813C}, SAMI \citepads[][]{2012MNRAS.421..872C}, MaNGA \citepads{2015ApJ...798....7B}, CALIFA \citepads{2016A&A...594A..36S}, or TIMER \citepads{2015A&A...584A..90G, 2019MNRAS.482..506G}.\looseness-2

Here we use IFS data to study the secular evolution of galaxies by analysing the kinematics of different stellar populations, as well as their relationship with parameters like dust content, stellar age and metallicity. In past studies, \citetads{1990MNRAS.245..350D} and \citetads{1992ApJ...400L..21B} quantified how absorption affects the rotation curves in the inner regions of late-type galaxies. These works were later supplemented with mainly Monte Carlo simulations for low surface brightness \citepads{2001ApJ...548..150M}, elliptical \citepads{2000MNRAS.313..153B,2000MNRAS.318..798B}, and disc galaxies \citepads{2003MNRAS.343.1081B}, and with studies using IFS data cubes leading to analytical methods for extracting the kinematics and the stellar population parameters \citepads[e.g.][]{2006MNRAS.365...74O, 2007IAUS..241..175C}. These methods, based on the fitting of IFS data in their full spectral range to single-age and single-metallicity stellar population (SSP) models \citepads{2003MNRAS.344.1000B, 1999ApJ...513..224V, 2010MNRAS.404.1639V}, have been successfully applied to several scenarios: the separation of counter-rotating populations in discs \citepads[][]{2013A&A...549A...3C, 2013MNRAS.428.1296J}, bulge-disc decomposition \citepads{2017MNRAS.465.2317J, 2017MNRAS.466.2024T, 2019MNRAS.485.1546T, 2019MNRAS.484.4298M, 2019MNRAS.488L..80M}, galaxy-halo separation \citepads{2018MNRAS.478.4255J}, or kinematically decoupled cores \citepads[KDC,][]{2018MNRAS.480.3215J}.\

\citet{etheses4179}\footnote{Available at Durham E-Theses \url{http://etheses.dur.ac.uk/4179/}} studied the kinematics of different stellar populations by assuming that different spectral regions are sensitive to stars of different ages. He extracted the line-of-sight velocity distribution parameters (\textit{v$_\star$}, \textit{$\sigma_\star$}, \textit{$h_3$}, and \textit{$h_4$}) for a young stellar population, traced by the H$\beta$ line, and an old one, traced by the Ca\,{\sc II} Triplet, and showed that the enhanced H$\beta$ absorption exhibited by elliptical galaxies may be produced by the presence of a disc formed by young stars.\

The aim of the present work is to quantify how the kinematics of young ($\lesssim 2$\,Gyr) and old stellar populations differ in the nuclear rings of a small sample of disc galaxies, and to explore relations with age, stellar metallicity, and extinction. We extract the line-of-sight velocity and velocity dispersion to characterise the kinematics of the young and old stellar populations, using the H$\beta$ line and the Ca\,{\sc II} Triplet, respectively. We compare our results with ages and metallicities obtained from the analysis of the data using the Galaxy IFU Spectroscopy Tool (GIST) pipeline \citepads[][]{2020arXiv200901856B}, and the extinction as obtained from the fitting of the full spectra. 

The paper is structured as follows. In Section~\ref{sec:sample} we present the galaxies we used for our study. Section~\ref{sec:methods} provides a detailed explanation of the methods and tools employed for the analysis of the data, and Section~\ref{sec:results} summarises the results obtained from our analysis. We discuss our results in Section~\ref{sec:discussion} and present a brief summary in Section~\ref{sec:conclusions}.

\begin{figure*}[!ht]
\centering
\includegraphics[width=\textwidth]{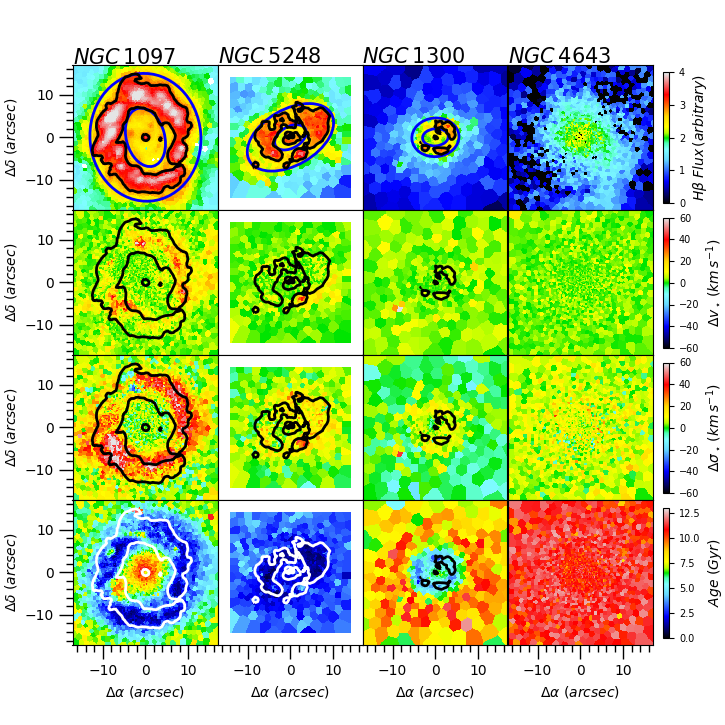}
\caption{Maps of kinematic and stellar population parameters for the four galaxies in our sample. (Top row) H$\beta$ emission-line fluxes in log scale. For NGC\,1097, NGC\,5248, and NGC\,1300 we overlay the flux level that defines the nuclear ring. These isophotes are also indicated in the other panels for reference, but not in NGC\,4643 since there is no evidence for a ring in this galaxy and it is used as a control galaxy. (Second and third rows) Maps of the difference in line-of-sight velocity and velocity dispersion between the two spectral ranges used in our study (i.e. H$\beta-$Ca\,{\sc II}). (Bottom row) Maps for the mean luminosity-weighted age from the analysis performed by \citetads[][]{2020arXiv200901856B}.}
\label{fig:fluvelage}
\end{figure*}

\begin{figure*}[!ht]
\centering
\includegraphics[width=\textwidth]{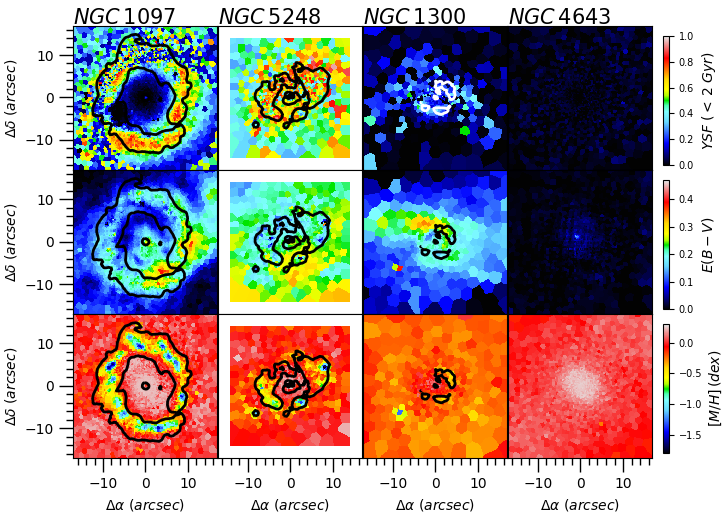}
\caption{Maps of dust extinction and stellar population parameters for the four galaxies in our sample. Each row (from top to bottom) presents: the fraction of young stars, colour excess $E(B-V)$, and metallicity [M/H] (see text for details). Isophotes as indicated in Fig.~\ref{fig:fluvelage}.}
\label{fig:frametcol}
\end{figure*}

\section{Sample}
\label{sec:sample}

We use data from the Multi-Unit Spectroscopic Explorer \citepads[MUSE;][]{2010SPIE.7735E..08B}, an integral field spectrograph installed on the Very Large Telescope (VLT) located at the Paranal Observatory in Chile. The spectral range is 4750\,\r{A} to 9300\,\r{A}, with a spectral resolution that ranges from 1700 for the blue wavelengths to 3400 for the red ones \citepads[][]{2017A&A...608A...1B, 2017A&A...608A...5G}. This instrument offers a $1\arcmin\,\times\,1\arcmin$ field of view with $0.2\arcsec$ per pixel in the Wide Field Mode. These capabilities allow the production of detailed maps in terms of spatial and velocity resolution, both crucial for the study of the kinematics of the inner structures of galaxies.\

We selected four galaxies from the Time Inference with MUSE in Extragalactic Rings (TIMER) sample \citepads{2019MNRAS.482..506G} of 24 barred nearby galaxies which present a wide range of star-forming rings. All TIMER galaxies are selected from the {\textit Spitzer} Survey of Stellar Structure in Galaxies \citepads[S$^4$G;][]{2010PASP..122.1397S} and are classified as strongly barred galaxies \citepads[SB,][]{2015ApJS..217...32B}. They also present high stellar masses, with values between $2.0\times10^{10}$\,M$_{\odot}$ and $1.7\times10^{11}$\,M$_{\odot}$ (as estimated in \citealtads{2013ApJ...771...59M}, \citealtads{2015ApJS..219....3M}, and \citealtads{2015ApJS..219....5Q}) and low to intermediate inclinations (below $60\degree$). The four selected galaxies were chosen to have a wide range of circumnuclear properties (e.g. the presence and size of a nuclear ring and the level of star formation).\

NGC\,1300 is an SB(rs)bc galaxy without nuclear activity \citepads{2004AJ....128.1124S}. It is an archetypal example of a strongly barred galaxy and has been used in many studies related to the dynamical modelling of galaxies \citepads[e.g.][]{1996A&A...313..733L, 1997A&A...317...36L, 2000A&A...361..841A}. While the presence of a nuclear ring is not evident given the smooth appearance of the stellar light in the inner regions \citepads[e.g.][]{1989ApJS...71..433P, 2000MNRAS.317..234P}, \citetads{2008AJ....135..479B} found evidence of a nuclear ring of 400\,pc radius \citepads[][]{2010MNRAS.402.2462C} traced by hot spots of Br$\gamma$ and [Fe\,{\sc ii}] emission lines. These lines are related to the photo-ionised gas around OB stars and supernova remnants respectively, thus tracing the regions where star formation activity may be taking place. 

NGC\,5248 is classified as an SAB(rs)bc galaxy in the NASA/IPAC Extragalactic Database (NED)\footnote{\url{http://ned.ipac.caltech.edu/}}, and it is a prime example of a grand-design spiral. There is a very active star-forming circumnuclear ring of 650\,pc radius \citepads[][]{2010MNRAS.402.2462C} populated by numerous young star clusters \citepads{1996AJ....111.2248M, 2001AJ....121.3048M}. \citetads{2001MNRAS.324..891L} and \citetads{2001AJ....121.3048M} detected a second small ring \citepads[of 150\,pc radius;][]{2010MNRAS.402.2462C} inside the first one using H$\alpha$ maps.\looseness-2

In addition to these two galaxies, we selected NGC\,1097 and NGC\,4643 as cases that illustrate better the extremes of high and low star formation activity, respectively, in the sample. \citetads{2019MNRAS.482..506G} also employed these two galaxies as examples of the wide range of properties  that can be found in the TIMER sample, performing an analysis of the stellar kinematics, age, metallicity and star formation history in the wavelength range between $4750$\,\r{A} and $5500$\,\r{A}. NGC\,1097 is an SB(s)b galaxy and like NGC\,5248 is a very well-studied grand-design spiral. It also harbours a 1\,kpc-radius \citepads[][]{2010MNRAS.402.2462C} nuclear ring which has been observed in several ranges of the electromagnetic spectrum revealing the presence of numerous stellar clusters  and a considerable amount of molecular gas \citepads[e.g.][]{2019MNRAS.485.3264P}. NGC\,4643 was added to our sample as a control galaxy. It is an SB(rs)0/a galaxy without a nuclear ring, which hosts little gas and in which the old component of the stellar population dominates the disc \citepads[][]{2019MNRAS.482..506G}.

\section{Analysis methods}
\label{sec:methods}

We aim to derive the stellar kinematics from different spectral regions and explore any differences observed in velocity and velocity dispersion, as well as possible correlations with key parameters. The first step in this analysis is to select the appropriate wavelength regions which may be sensitive to, at least, extreme stellar populations (e.g. young and old). Following \citetads{etheses4179}, we used the H$\beta$ and Mg$b$ region ($4750-5500$\,\r{A}), and the Ca\,{\sc II} Triplet region ($8498-8950$\,\r{A}) as proxies for young and old populations, respectively.

H$\beta$ is particularly strong in absorption for stars with spectral type A. We thus assume that the kinematics derived from the spectral fitting around this line may be a good tracer of the kinematics of a relatively young stellar population. On the other hand, the Ca\,II Triplet is most prominent in absorption in G, K, and M type stars, tracing the old stellar population. 

In our analysis we make use of the stellar population analysis from \citetads[][]{2020arXiv200901856B} obtained with the GIST pipeline\footnote{\url{http://ascl.net/1907.025}} \citepads{2019A&A...628A.117B} that was applied to the entire TIMER sample. This analysis was performed imposing a level of Voronoi binning with a target signal-to-noise ratio of 100 per spectral pixel. The stellar kinematics ($v_\star$, $\sigma_\star$, h$_3$, h$_4$) and the stellar population parameters (e.g. luminosity-weighted age, metallicity [M/H], and [$\alpha$/Fe]) were extracted from a specific wavelength range (from $4800-5800$\,\r{A}). The MILES stellar population models \citepads{2011A&A...532A..95F, 2015MNRAS.449.1177V} were used as templates for the fitting.

Since we are interested in the kinematics of the two distinct spectral regions, we re-run the GIST pipeline for the kinematics extraction for each of them separately. In this second analysis we masked the emission lines that can affect the fitting process, like the H$\beta$ line itself. In order to make our results easily comparable with those computed in the $4800-5800$\,\r{A}\ wavelength range, we imposed the same Voronoi binning. While ideally it would be desirable to also extract the stellar population parameters from each spectral range, the Ca\,{\sc II} Triplet region by itself is not sensitive enough to obtain reliable ages, metallicities, and [$\alpha$/Fe] values. This is why we rely on the average values presented in \citetads[][]{2020arXiv200901856B}, which are sensitive to the presence of young stellar populations. Besides the fitting in each spectral region, we performed an additional fitting for the whole spectral range in order to extract the colour excess $E(B-V)$ from the binned spectra. In this second run we used the E-MILES models \citepads[][]{2016MNRAS.463.3409V} as templates, which are identical to the MILES models in the $4800-5800$\,\r{A}\ region, but allow us to extend the analysis to the Ca\,{\sc II} Triplet.

Three of the galaxies in our sample present well-defined nuclear rings characterised by the presence of young stars. NGC\,4643 does not, is dominated by old stellar populations, and is used here as a control galaxy. Since we want to study specific relations between the kinematic and the stellar population parameters in different regimes, we have identified different regions of the galaxies with distinct properties. We use H$\beta$ flux maps to help us define the nuclear ring-dominated areas. Regions outside those limits are denoted as the {\it disc}, while the area inside the ring is denoted as the {\it nucleus} (see Fig.~\ref{fig:fluvelage} and \ref{fig:frametcol}). We note that we use only a portion of the MUSE field of view (FoV) for this analysis. This selection is done with the purpose of avoiding the noisy spaxels at the edges of the cubes and possible artifacts. The selected portion of the MUSE FoV covers an area of $17\arcsec\times17\arcsec$ in all the galaxies in the sample, with the exception of NGC\,5248, where the area is slightly smaller ($14\arcsec\times14\arcsec$).

\section{Results}
\label{sec:results}

We present our results for the three regions identified for the galaxies of our sample: disc, ring, and nucleus. The results are summarised in Table~\ref{tab:results} and include difference in line-of-sight velocity, difference in velocity dispersion, age, fraction of young stars, colour excess, metallicity, and colour.

\subsection{Kinematic differences}
\label{sec:kinematics}

The main result of our study is the kinematic difference between the two selected spectral ranges. The four galaxies selected from the TIMER sample illustrate scenarios where a young stellar population is present to different extents in the inner regions of the galaxies: from the extremely young inner disc of NGC\,5248 to the lack of evidence of young stellar populations in the inner regions of NGC\,4643. We now seek to confirm whether the young stars leave an imprint in the stellar kinematics, especially in the rings. Throughout the paper we refer to the differences in line-of-sight velocity between the two spectral ranges (young minus old) as $\Delta v_\star$, calculated as the absolute difference of the values of the estimated velocities for the H$\beta$ and Ca\,{\sc II} Triplet lines.

The maps in Fig.~\ref{fig:fluvelage} present a fairly homogeneous and smooth appearance with the exception of NGC\,1097 and some small regions in NGC\,5248. In NGC\,1097 we can clearly identify the shape of the ring, traced mostly by regions where the \deltaV\ values are high. This ring spans a wide range of \deltaV\ values, up to $70\,{\rm km\,s^{-1}}$, while in the disc and nucleus the maximum variation is about $25\,{\rm km\,s^{-1}}$. These differences are somewhat milder in the other three galaxies in our sample with mean values close to zero and a typical standard deviation of $5\,{\rm km\,s^{-1}}$. There are a few exceptions to this behaviour in specific locations in the disc of NGC\,1300 and the ring of NGC\,5248.

The presence of the rings stands out more when inspecting the differences in velocity dispersion (\deltaS\ , also computed as young minus old). This is most notable in NGC\,1097 with mean \deltaS\ \,values of $24\,{\rm km\,s^{-1}}$, well above the mean values observed in its disc and nucleus ($\sim\,10\,{\rm km\,s^{-1}}$ and $\sim\,5\,{\rm km\,s^{-1}}$, respectively). For the other three galaxies, the scenario is analogous to what we observe for \deltaV\ . It is worth noting the slightly positive mean overall value of \deltaS\ observed in NGC\,4643 ($8\,{\rm km\,s^{-1}}$). Given the lack of young stellar populations the expectation is to find \deltaS\ \,differences close to zero. If this was a problem due to the adopted spectral resolution as a function of wavelength in our analysis, one would expect a similar offset in equally old regions of other galaxies. This is however not observed when inspecting the disc dominated regions of NGC\,1300, with populations similar to those in NGC\,4346 but mean \deltaS\ \,of $1\,{\rm km\,s^{-1}}$. With uncertainty levels for \deltaS\ \,in our measurements of $9\,{\rm km\,s^{-1}}$ (see \citealtads{2019A&A...623A..19P}) it is difficult to draw more firm conclusions.

\subsection{Age}
\label{sec:age}

Age is an excellent tracer of the presence of the rings in our sample. As expected, NGC\,4643 is very different from the other three galaxies in the sample. In this galaxy there is no ring structure in the age map, which appears rather homogeneous. In NGC\,4643 we find a rather old stellar population with a mean age around $11$\,Gyr with a small standard deviation of $0.7$\,Gyr. 

In NGC\,1097 and NGC\,5248 we do not only find a well defined ring-like structure, but also relatively young stellar populations within them. This is especially noticeable in the ring of NGC\,5248. In the two rings we estimate an average age below $4$\,Gyr ($3.5$\,Gyr for NGC\,1097 and $1.8$\,Gyr for NGC\,5248). It is remarkable that in NGC\,5248 we find a relatively small range of ages (from $0.5$\,Gyr to $5.0$\,Gyr) while in the ring of NGC\,1097 it is much larger (from $0.8$\,Gyr to $10.7$\,Gyr), which clearly affects the obtained mean value. On the other hand, the ring of NGC\,1300 seems to be dominated by an intermediate-age stellar population, with a mean age of $4.6$\,Gyr. The ages in all these rings correspond reasonably well with the ellipses we used to define them (see Sect.~\ref{sec:methods} for details). 

The other two regions defined in these galaxies display different behaviours. NGC\,1097 has a disc with a stellar population of intermediate age ($\simeq 6$\,Gyr) and a central region dominated by older stars (with mean age over $8$\,Gyr). In NGC\,1300 the disc has an average age of $8.3$\,Gyr, while the region inside the ring has a slightly younger mean age of $5.9$\,Gyr with a smaller dispersion ($0.7$\,Gyr for the central region and $1.6$\,Gyr for the disc). The whole disc of NGC\,5248 is populated by young stars. The three defined regions are dominated by young stars with mean ages below $3$\,Gyr and a small dispersion which never exceeds $0.7$\,Gyr. All the age measurements agree well with the typical age uncertainties: around $0.5$\,Gyr and $3$\,Gyr for the young and old stellar population, respectively \citepads[][]{2019A&A...623A..19P}.

\subsection{Fraction of young stars}
\label{sec:ysf}

The fraction of young stars is recovered from the light-weighted results obtained by \citetads[][]{2020arXiv200901856B}. We sum the weights applied during the fitting process to all the templates corresponding to stellar populations below a certain age, considering the models with ages below $2$\,Gyr as those tracing the young stellar populations. We checked that lowering this value to $1$\,Gyr gave consistent results. 

The maps for the fraction of young stars (top row in Fig.~\ref{fig:frametcol}) complete the information given by the age maps. NGC\,4643 is characterised by a very homogeneous distribution of low values of the fraction of young stars, with a mean value close to zero and never exceeding $0.13$. The other three galaxies illustrate different scenarios for the presence of a well defined ring. In the case of NGC\,1097 the ring exhibits large regions dominated by high fractions of young stars, with a mean value of $0.44$ with a dispersion of $0.22$. In the disc we find similar values (mean value of $0.33$ and a standard deviation of $0.18$), while in the nucleus we find extremely low fractions with a mean value of $0.10$ and a standard deviation of $0.10$.

NGC\,5248 shows high fractions of young stars in all the regions. The ring is recognisable by a greater presence of bins with high fractions of young stars, resulting in a slightly higher mean value, $0.63$, when compared with the other two regions ($0.53$ and $0.55$ for the disc and nucleus, respectively). In all the three regions the dispersion of values is small, around $0.15$. In contrast, NGC\,1300 exhibits rather low fractions in all the regions, with mean values of $0.09$, $0.23$, and $0.019$ in disc, ring, and nucleus, respectively, and dispersion around $0.10$. The typical uncertainty in our measurements of the fraction of young stars is $0.13$.

\subsection{Extinction}
\label{sec:ebv}

The maps of extinction, characterised by the colour excess $E(B-V)$ (second row in Fig.~\ref{fig:frametcol}), reveal a little more of the structure of the galaxies. These maps replicate and emphasise the information given by the colour maps in \citetads{2019MNRAS.482..506G}. For NGC\,1097 the $E(B-V)$ map shows not only the presence of dust lanes in the ring, but also along the leading edge of the bar. In this galaxy we find a wide range of values of the colour excess in the three regions, resulting in discrete mean values in each one of them ($0.08$ in the disc, $0.16$ in the ring, and $0.14$ in the nucleus, all with low standard deviations around $0.05$). In the map for NGC\,1300 we also recognise these structures traced by higher values of $E(B-V)$ when compared with those for the rest of the galaxy. Regarding the mean values in the three regions, NGC\,1300 exhibits higher values compared to those in NGC\,1097: $0.14$ for the disc, $0.21$ for the ring, and $0.20$ for the nucleus, all with dispersion below $0.07$. NGC\,5248 is slightly different, with in general higher values of the colour excess. This gives us the image of a dusty galaxy, with plenty of star-forming regions. The ring is traced by alternating hot-spots of low and high values of the extinction, like in NGC\,1097. For this galaxy the values in the three regions are very similar: $0.23$ for the disc, $0.20$ for the ring, and $0.218$ for the nucleus. The $E(B-V)$ map for NGC\,4643 follows a similar pattern to the colour map obtained for this galaxy, presenting a patchy distribution of very low values (with a mean value around zero). All our measurements of the colour excess are expressed with an uncertainty of $0.05$.

\begin{figure*}[!ht]
\centering
\includegraphics[width=\textwidth]{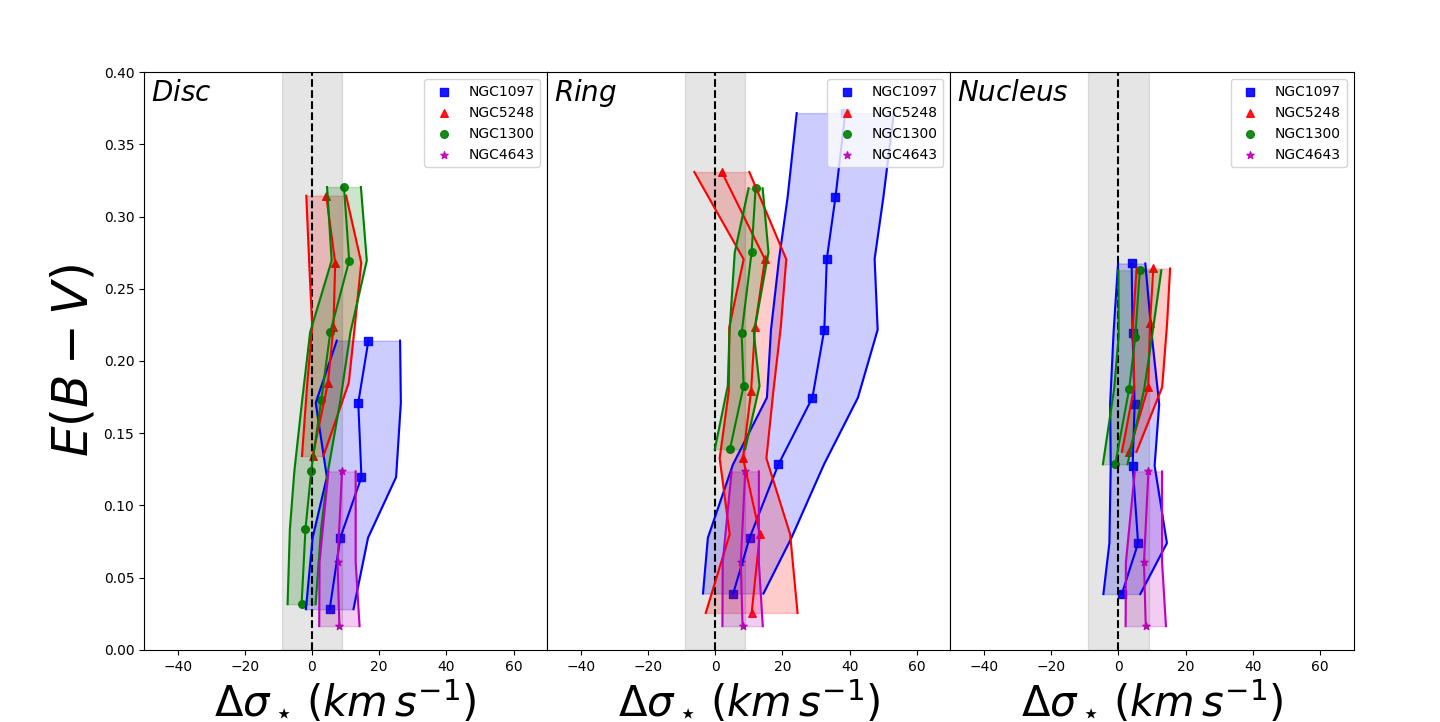}
\caption{Observed values of $\Delta\sigma_\star$ against colour excess $E(B-V)$ in the three selected regions for the four galaxies in the sample. We represent the mean values of $\Delta\sigma_\star$ estimated in equally spaced age bins. We also represent the $1\,\sigma$ uncertainty interval for these mean values as colour-shaded regions. The typical error in the $\Delta\sigma_\star$ estimate is represented as the grey-shaded area. In general we find vertical distributions of the mean values of $\Delta\sigma_\star$. The only exception is the ring of NGC\,1097 where we find increasing values of $\Delta\sigma_\star$ when $E(B-V)$ increases.}
\label{fig:scat_dsig-ebv}
\end{figure*}

\subsection{Metallicity}
\label{sec:metallicity}

The metallicity also reproduces quite well the ring-like structure in all the galaxies of our sample (see the third row in Fig.~\ref{fig:frametcol}). The only exception is NGC\,4643 in which we find some structure, corresponding to the presence of a nuclear disc, but not a ring, with higher values in the central few hundred parsecs of the galaxy. We also see a band of intermediate and high values that goes from the south-west to the north-east part of the disc and which corresponds to the bar. In general we observe a rather homogeneous distribution around the mean value of $-0.25$\,dex. 

In NGC\,1097 there is a clear match between the H$\beta$ amplitude line contours used to define the ring and the regions with lower metallicities (ranging from $-1.7$ to $0.24$\,dex with an average value of $-0.33$). In the central region and the rest of the disc we find similar mean values of $0.15$ and $0.04$, respectively. 

In NGC\,1300 we report a similar match between regions with low metallicity and the isophotes obtained from the H$\beta$ line amplitude map. For this galaxy we observe lower metallicity throughout the disc compared to the other galaxies in the sample. It is noticeable that for this galaxy the H$\beta$ line isophotes used do not cover an important part of the ring. In these areas the metallicity reaches values similar to the central part of the galaxy (around $0$). The mean value of the metallicity in the three regions is similar: $-0.22$\,dex for the disc, $-0.18$ for the ring, and zero for the nucleus. 

NGC\,5248 also shows a correspondence between the peaks in the H$\beta$ line amplitude map and the low metallicity regions, and, like in the case of NGC\,1300, in the areas where the isophote is disrupted the metallicity reaches higher values, similar to the disc and the nuclear region. The mean values in the three regions defined ($-0.09$ in the disc, $-0.21$ in the ring, and $0.15$ in the nucleus) are quite similar and also to the standard deviations in each region ($0.15$, $0.30$, and $0.12$\,dex, respectively). The typical uncertainty in the measurements of the metallicity in all the galaxies in our sample is around $0.10$\,dex \citepads[][]{2019A&A...623A..19P}.\

\begin{figure*}[!ht]
\centering
\includegraphics[width=\textwidth]{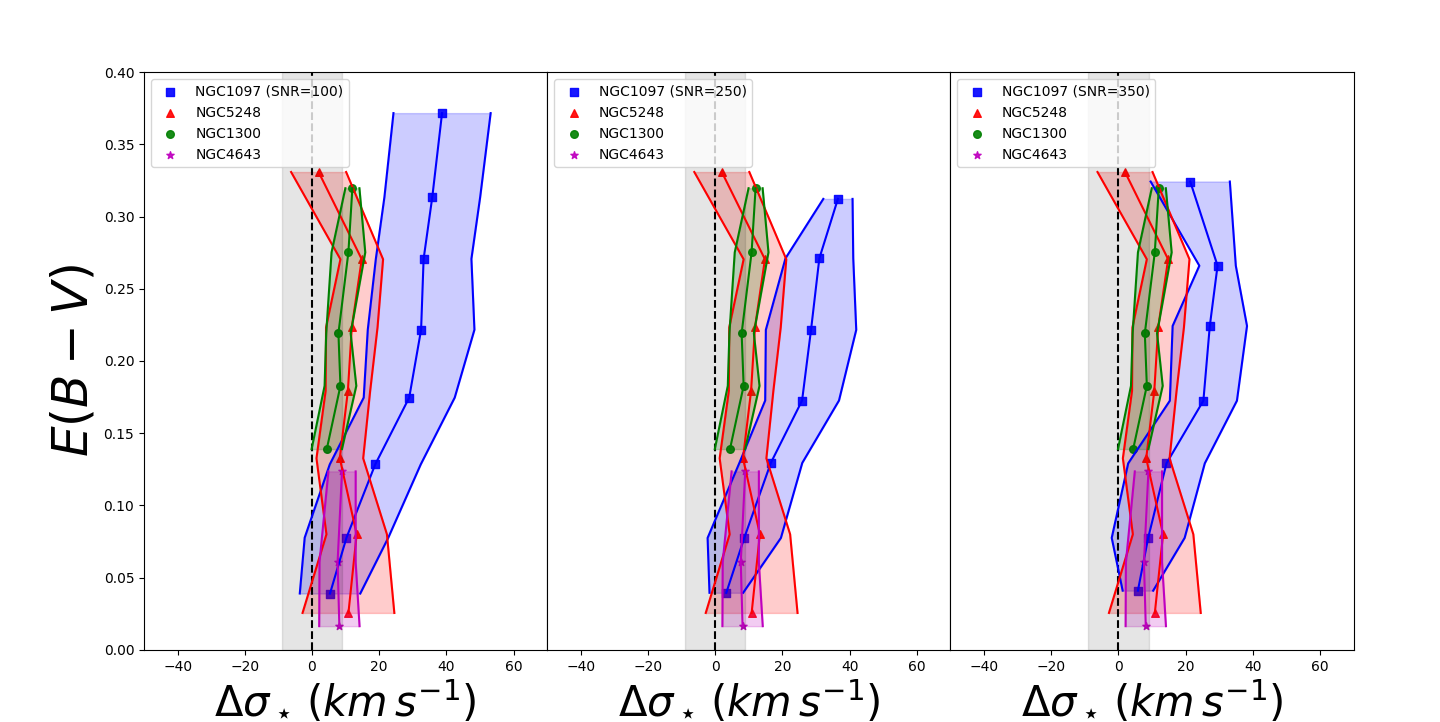}
\caption{Changes in $\Delta\sigma_\star$ with the $E(B-V)$ distribution in the ring in NGC\,1097. We plot the results for three different target SNR for the GIST pipeline analysis (the blue series): $100$ (left panel), $250$ (central panel), and $350$ (right panel). We can observe how the highest values of the colour excess disappear when we increase the SNR. At the same time the highest values of $\Delta\sigma_\star$ start to decrease. At SNR=350 we cannot clearly discern the trend we observe for SNR=100 and the distribution for the ring of NGC\,1097 is similar to that for that of NGC\,5248.}
\label{fig:scat_dsig-ebv_snr}
\end{figure*}

\section{Discussion}
\label{sec:discussion}

In light of the results obtained, we now wish to confirm which parameters depend on or influence the kinematics estimated in the spectral regions around the H$\beta$ and Ca\,{\sc II} Triplet lines. As mentioned above, we operate under the hypothesis that the kinematics derived using these spectral regions are sensitive to different stellar populations. We study the possible trends of the difference between the kinematics of young and old stellar populations with the various parameters extracted. We pay special attention to the behaviour of the rings, but also inspect trends in the disc and nucleus components.

\begin{figure}[!ht]
\centering
\includegraphics[scale=0.5]{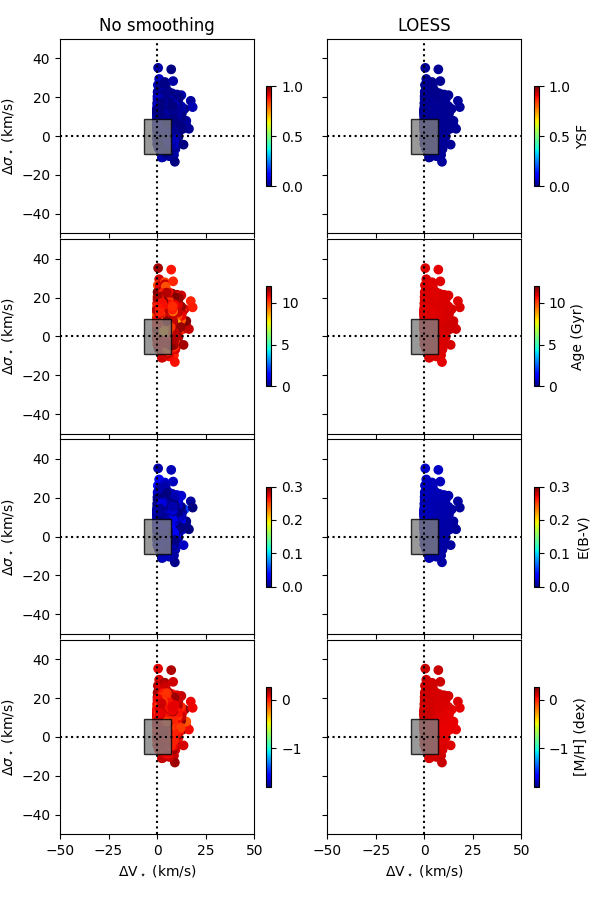}
\caption{$\Delta\sigma_\star$ against $\Delta v_\star$ with colour-coded values of the stellar population parameters (from top to bottom: fraction of young stars, mean age, colour excess, and metallicity) for NGC\,4643. On the left side we represent the data with no smoothing and on the right side the \textsc{loess}-smoothed data. The grey rectangle represents the typical error in $\Delta v_\star$ ($7$\,\kms) and $\Delta\sigma_\star$ (9\,\kms). The uncertainties introduced by the smoothing are: $\pm 0.019$ for the YSF, $\pm 0.7$\,Gyr for the mean age, $\pm 0.016$ for $E(B-V)$, and $\pm 0.08$\,dex for the metallicity.}
\label{fig:ppxf_4643}
\end{figure}

\begin{figure}[!ht]
\centering
\includegraphics[scale=0.5]{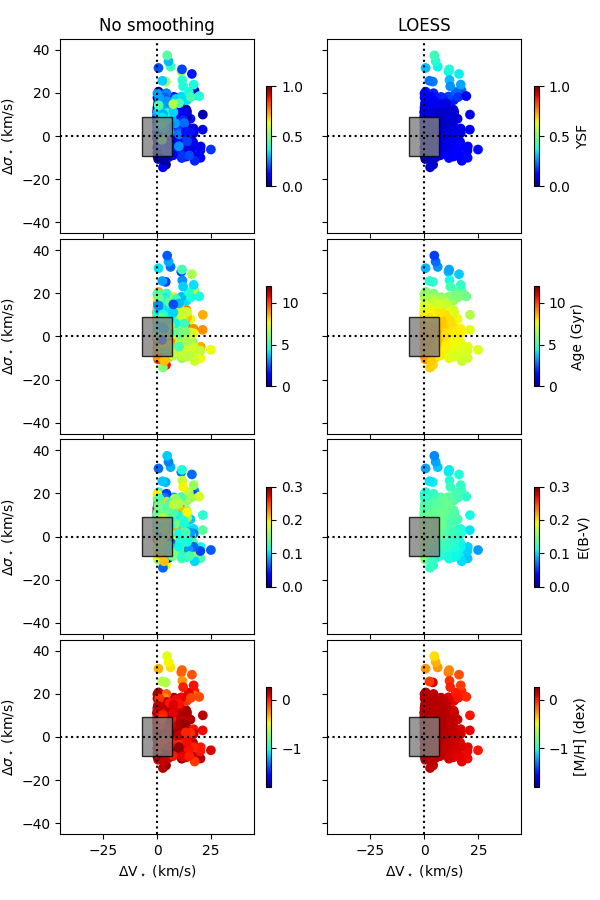}
\caption{As Fig.~5, now for the nucleus of NGC\,1097. The uncertainties introduced by the smoothing are: $\pm 0.10$ for the YSF, $\pm 1.7$\,Gyr for the mean age, $\pm 0.06$ for E(B$-$V), and $\pm 0.10$\,dex for the metallicity.}
\label{fig:ppxf_nuc_1097}
\end{figure}

\subsection{Trends of the kinematic differences with stellar population properties}

We start to explore the relation between the kinematic (velocity and velocity dispersion) differences of the two spectral regions studied and the stellar population properties extracted from our analysis. We first consider the nuclear ring regions.

Figure~\ref{fig:scat_dsig-ebv} and the corresponding figures in Appendix~\ref{append: kin_vs_prop} show the differences in velocity (\deltaV\ ) and velocity dispersion (\deltaS\ ) plotted against the values of mean age, fraction of young stars, metallicity, and colour excess. In order to make our analysis of the possible trends easier, we do not directly represent the whole clouds of points. We separate the results in each region of the galaxies in equally spaced bins for each parameter and estimate the mean value of the kinematical differences in each one of them. Additionally, for robustness, we ignore any bins with less than three elements. We study the significance of the mean value in each bin by comparing the $1\,\sigma$ confidence interval with the typical error in our estimations (calculated as the mean standard deviation of the errors of the difference in velocity and velocity dispersion, giving values of $7\,{\rm km\,s^{-1}}$ and $9\,{\rm km\,s^{-1}}$, respectively).

Examining the plots for \deltaV\ (see Fig.~\ref{fig:scat_dv-age}, Fig.~\ref{fig:scat_dv-ysf}, Fig.~\ref{fig:scat_dv-metal}, Fig.~\ref{fig:scat_dv-ebv}), we find no significant trends in any of the regions, with the exception of \deltaV\ \,vs [M/H] where there is a tentative trend with metallicity in NGC\,1097 and NGC\,5248, the galaxies with the largest rings. The observed trend is such that high \deltaV\ \,values are found for lower metallicities. There is some evidence that the metal-poor stars in the ring of NGC\,1097 are due to an interaction event with a nearby galaxy which has rejuvenated the ring \citep[see][for details]{2020arXiv200901856B}, a phenomenon which could explain this trend.

Regarding \deltaS\,, contrary to what one may expect, we do not observe clear trends with age or with the fraction of young stars (Fig.~\ref{fig:scat_dsig-age} and Fig.~\ref{fig:scat_dsig-ysf}, respectively). The strongest relation seems to be with E(B$-$V). In Fig.~\ref{fig:scat_dsig-ebv} we show that there is a trend of $\Delta\sigma_\star$ with the colour excess, at least in the case of the ring of NGC\,1097. There we find a clear trend of increasing values of $\Delta\sigma_\star$ with increasing colour excess. It is especially remarkable for the highest values of E(B$-$V), where we observe high values of the mean difference of velocity dispersion. These values are equivalent to several times the typical error. In the other galaxies and the other regions of NGC\,1097 the mean values of $\Delta\sigma_\star$ in the colour excess bins are similar to the error. It is noticeable that the $1\,\sigma$ confidence intervals for the values of $\Delta\sigma_\star$ in the ring of NGC\,1097 are slightly higher than the typical error. Nevertheless, we consider the \deltaS\ \,measurements in the ring as robust as those in the other regions of NGC\,1097 and the other galaxies in the sample. For the remaining parameters, we do not find clear trends (the relevant plots can be found in the appendix).\looseness-2

We do not find any significant relation of the kinematic differences with the extracted stellar population parameters in the cases of NGC\,1300 and NGC\,5248. We therefore will focus our attention on NGC\,1097 and NGC\,4643 only during the discussion, as they represent the two extreme cases of a very active ring on the one hand and a quiescent galaxy that does not exhibit any relevant kinematic difference on the other. The cases of NGC\,1300 and NGC\,5248 are presented in the appendix section instead.

Why do we see the trend in the ring in NGC\,1097 and not in those in the other galaxies? We suggest that this is due to a beam-smearing effect, more pronounced in NGC\,1300 and NGC\,5248 than in NGC\,1097. To confirm this, we re-ran the analysis of the data of NGC\,1097 using the GIST pipeline at two different signal-to-noise ratios (SNR): $250$ and $350$. This change in the target SNR produces bigger Voronoi bin sizes and thus poorer spatial resolution. In the case of NGC\,1097, for the considered field of view, at a SNR=$100$ we obtain $3609$ bins, at a SNR=$250$, $540$ bins, and at a SNR=$350$, $264$ bins. In contrast, for NGC\,1300 at SNR=$100$ we have $537$ bins. Considering the distances for these two galaxies and a target SNR=$100$, the average size of a bin located in the ring of NGC\,1300 is about $39{\rm\,pc}$, compared to around $22{\rm\,pc}$ in NGC\,1097 ($74{\rm\,pc}$ at SNR$=350$). The effect of the change of the bin size on the ring of NGC\,1097 can be seen in Fig~\ref{fig:scat_dsig-ebv_snr}. There we show that increasing the bin size is translated into a loss of the higher values of the colour excess and a decrease in the higher values of $\Delta\sigma_\star$ of about $10\,{\rm km\,s^{-1}}$. We further discuss this in Appendix~\ref{append: kin_1300_5248}.\looseness-2

\begin{figure}[!ht]
\centering
\includegraphics[scale=0.5]{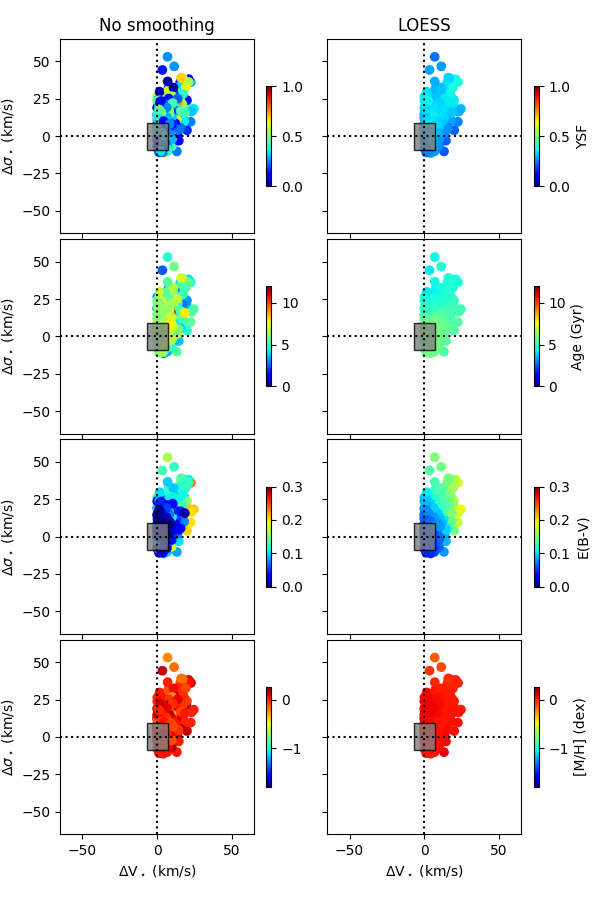}
\caption{As Fig.~5, now for the disk of NGC\,1097. The uncertainties introduced by the smoothing are: $\pm 0.17$ for the YSF, $\pm 1.4$\,Gyr for the mean age, $\pm 0.04$ for E(B$-$V), and $\pm 0.010$\,dex for the metallicity.}
\label{fig:ppxf_out_1097}
\end{figure}

\begin{figure}[!ht]
\centering
\includegraphics[scale=0.5]{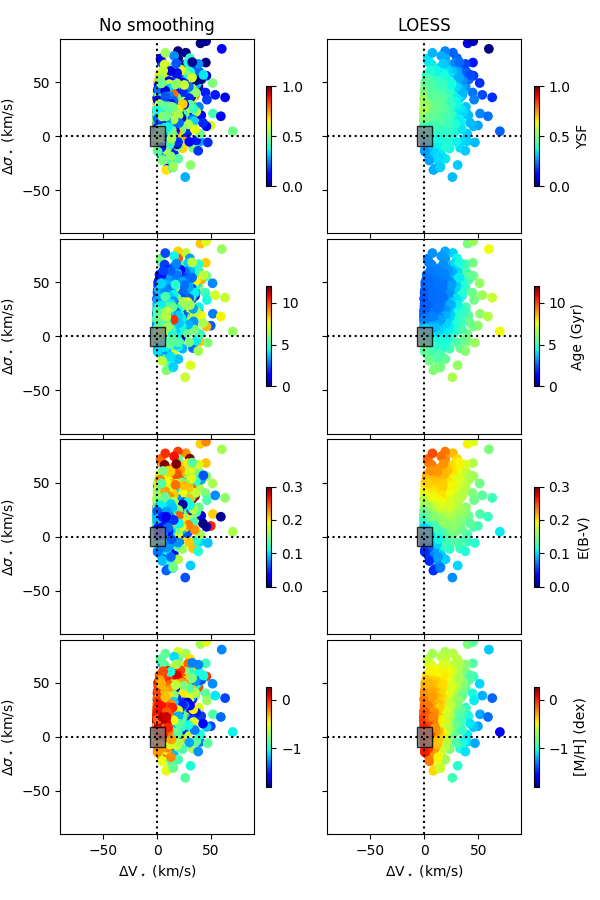}
\caption{As Fig.~5, now for the nuclear ring of NGC\,1097. The uncertainties introduced by the smoothing are: $\pm 0.20$ for the YSF, $\pm 1.6$\,Gyr for the mean age, $\pm 0.05$ for E(B$-$V), and $\pm 0.36$\,dex for the metallicity.}
\label{fig:ppxf_ring_1097}
\end{figure}

The main inconvenience with the way we present the trends in Figures~\ref{fig:scat_dsig-ebv} and \ref{fig:scat_dsig-ebv_snr} is that the most extreme values get diluted when averaging in the bins in which we separate our results. An alternative method to present our results is to confront \deltaV\ \,and \deltaS\ \,and colour-code the individual Voronoi bin values with different properties. The problem with this representation of the complete set of data is that for each range of values of the stellar population parameters we have a relatively large dispersion of the values of the kinematic differences. To aid the eye in identifying the main trends we have applied some smoothing using the \textsc{loess}\footnote{http://purl.org/cappellari/idl} package. This software is able to recover the trends in noisy data by applying a two-dimensional Locally Weighted Regression method \citepads{Clev:Devl:1988}. Results are presented in Figs.~\ref{fig:ppxf_4643}, \ref{fig:ppxf_nuc_1097}, \ref{fig:ppxf_out_1097}, and \ref{fig:ppxf_ring_1097} for the whole data set of NGC\,4643 and the three regions of NGC\,1097. In order to illustrate the changes introduced by the smoothing process, in Appendix~\ref{append: kin_1300_5248} we represent the original data in the left panel of each plot. We also account for the uncertainties introduced by the smoothing, estimating them as the standard deviation of the difference between the original and the smoothed data (as indicated in the caption of the figures).

NGC\,4643 (see Fig.~\ref{fig:ppxf_4643}) does not show any trend in the plots. The kinematic differences are clustered around low values, just a few times larger than the typical uncertainties. Regarding the values of the stellar population parameters, the dynamical ranges are small, with differences between the maximum and minimum values in general smaller than the estimated uncertainty.

The nucleus of NGC\,1097 (see Fig.~\ref{fig:ppxf_nuc_1097}) presents a similar case to that of NGC\,4643, spanning similar ranges of values. Its stellar population parameters show also relatively small dynamical ranges. There are some indications of weak trends, particularly of \deltaV\ \,with mean age. In the case of the disc (see Fig.~\ref{fig:ppxf_out_1097}) we find higher values of \deltaS\ \,, but low dynamical ranges of the parameters. There is a weak trend of \deltaV\ with colour excess, but the range of values of E(B$-$V) ($0.04-0.2$) is not large enough to be formally significant.

The ring of NGC\,1097 (Fig.~\ref{fig:ppxf_ring_1097}) exhibits larger kinematic differences between the two stellar populations. In general, all the stellar population parameters span wider ranges of values and trends of \deltaS\ \,with mean age and $E(B-V)$ are found. In the case of the trend with age there are indications of increasing \deltaS\ \,with younger ages, but the range of mean ages spanned in this region ($2-7$\,Gyr) is not large enough to confirm the trend. The picture with colour excess is quite different. We see a clear increase of \deltaS\ \,with $E(B-V)$, which spans a wide dynamic range ($0.06-0.27$). This results suggests that we may be probing different set of stars along the line-of-sight in the two spectral ranges, and this will contribute to the elevated \deltaS\ \,values.\newline

\subsection{Kinematic simulations}
\label{sec:sims}

When interpreting the results above, it is important to establish the expected level of kinematic differences caused by distinct populations alone (i.e. not influenced by the presence of dust). For that purpose, we have created a set of mock spectra made of young and old stellar populations with a wide range of kinematic properties. We designed the experiment to reproduce as closely as possible all the relevant details of the MUSE data, from the ranges used to the spectral resolution and sampling. We generated $\sim$48000 test cases using the E-MILES library as the reference with the following input parameters:

\begin{itemize}
   \item Fraction of young stars: [10$^{-2}$,10$^0$] (21 steps log-spaced)
   \item Age$_\mathrm{young}$: [0.1,0.4,1.0] (Gyr)
   \item Age$_\mathrm{old}$: 12.0 (Gyr)
   \item $\mathrm{[M/H]}_\mathrm{young}$: 0.06 dex
   \item $\mathrm{[M/H]}_\mathrm{old}$: 0.06 dex
   \item V$_\mathrm{young}$: [-100.0,100.0] (in steps of 25\,\kms)
   \item V$_\mathrm{old}$: [-100.0,100.0] (in steps of 25\,\kms)
   \item $\sigma_\mathrm{young}$: [50.0,150.0] (in steps of 50\,\kms)
   \item $\sigma_\mathrm{old}$: [150.0,250.0] (in steps of 50\,\kms)
\end{itemize}

\begin{figure}[!ht]
   \centering
   \includegraphics[width=\linewidth]{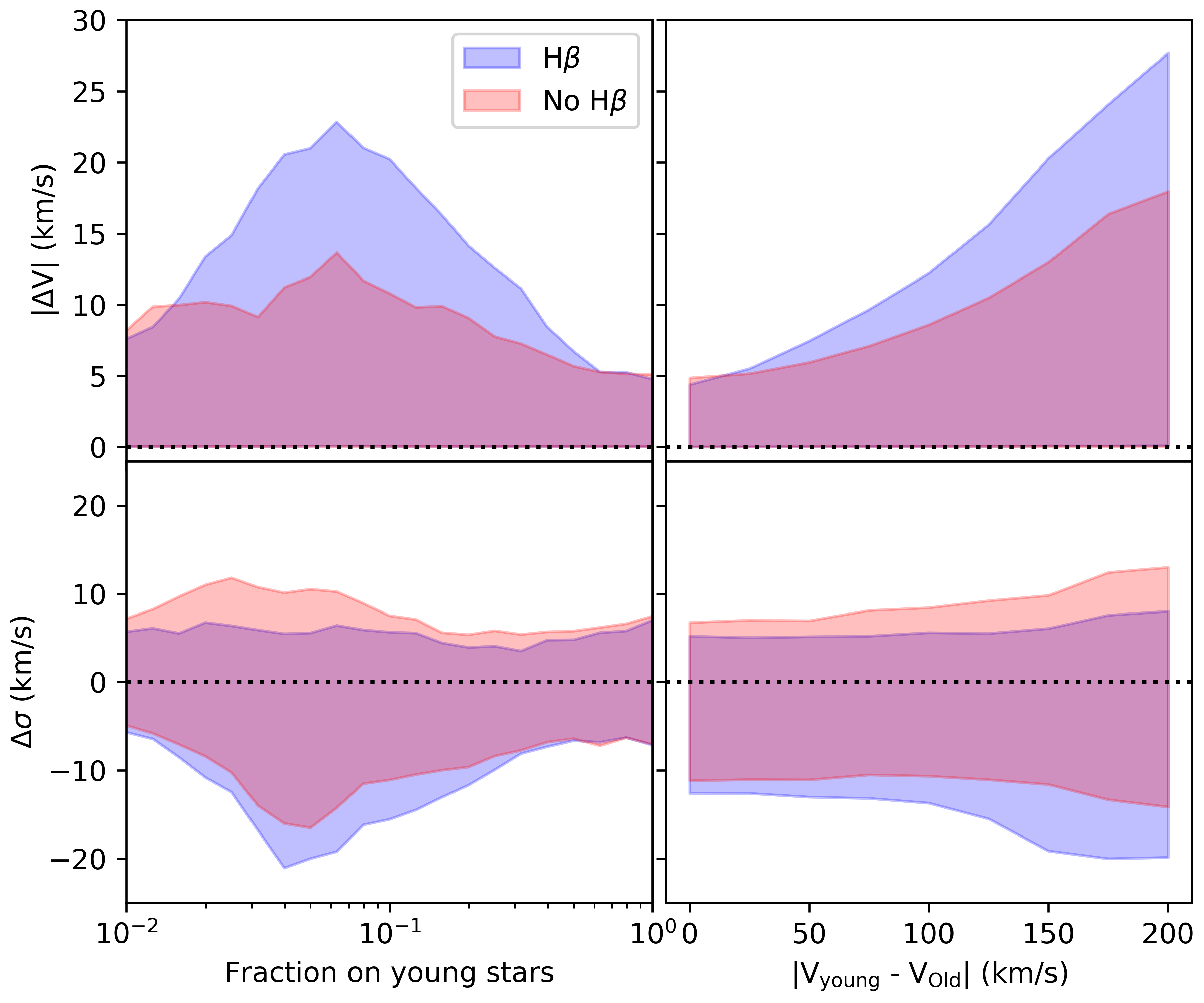}
   \caption{Kinematic differences between the H$\beta$ and Ca\,{\sc II} triplet regions extracted from our simulations. Top panel shows the differences in velocity and velocity dispersion as a function of: (1) fraction of young stars and (2) input velocity difference between populations. Shaded areas mark the $1\%-99\%$ percentiles sampled by our simulations when the H$\beta$ line is included (blue) or not (red) in the spectral fit.}
   \label{fig:sims}
\end{figure}

For consistency, we fitted the data using the GIST pipeline with the same configurations as used with the real data. This included the masking of emission line-dominated regions, even though emission lines were not actually included in the mock spectra. Given the prominence of the H$\beta$ line in the 4750$-$5500\,\r{A} region, we also performed a separate run without masking any parts of the spectra. The most relevant results of this experiment are presented in Fig.~\ref{fig:sims}. Rather than presenting all the individual measurements, which are heavily dependent on our choice of input parameters for the simulations, we present the areas covering the 1\%$-$99\%\ percentiles of the output distributions as a function of the input parameter. They provide a good estimate of the range of possible solutions. These regions are denoted by the shaded areas in the different panels. The red areas correspond to the fits masking the H$\beta$ line, while the blue ones are for fits where no pixels have been excluded.

The first striking, but perhaps not surprising result of our simulations, is the lower sensitivity to differentiate between populations when the H$\beta$ line is masked (i.e. as we find with real data). These differences are most notable in the velocity, and have a smaller effect in the velocity dispersion. This effect may explain the lack of trends observed with fraction of young stars in our MUSE data for all galaxies. The spread of most of the (red) solutions are also not much bigger than the typical uncertainty of our measurements (7\,\kms). Furthermore, the most extreme differences in velocity of $\sim$17\,\kms\ are for situations where the differences in velocity between the two populations is of the order of 200\,\kms, which is highly unlikely to be the case in our galaxies.

Another interesting feature in Fig.~\ref{fig:sims} is the dependence of the kinematic differences on the fraction of young stars. When this fraction is either too small or too large, the resulting spectra are so dominated by old or young stars respectively that the differences are within the uncertainties of our measurements, and one effectively has a spectrum of a single stellar population. In our specific tests the largest differences are observed for intermediate fractions of young stars of around $5$\%, and again only for extreme differences in velocity of the input populations.

Our experiments confirm that with our particular setup for the kinematic extraction (i.e. masking the H$\beta$ line), we should not observe strong trends with the fraction of young stars in the galaxies. The observed differences must thus be caused by dust extinction. This is an aspect that we cannot simulate without performing proper radiative transfer propagation to generate the input model spectra, which is well beyond the scope of this paper. 

The most striking fact about our results is that we have positive values of the difference of velocity dispersion $\Delta \sigma_\star = \sigma_{\rm H\beta} - \sigma_{\rm Ca\,II}$. This would mean, if we assume that the H$\beta$ line is dominated by the young stars and the Ca Triplet by older ones, that we are observing a young stellar population with a higher velocity dispersion. This seems to contradict the intuitive idea of young stellar populations having a more ordered motion than the old stellar populations. The simulations above indicate that, at the spectral resolution of our data, differences in stellar populations is not the cause of the observed kinematic differences.

\subsection{Asymmetric drift}
\label{sec:ad}

The asymmetric drift $v_a$ is one of the possible sources of an enhanced velocity dispersion in the kinematics traced by the H$\beta$ line spectral region. It can be defined as the difference between the velocity of a hypothetical ensemble of stars $v_c$, which moves around the centre of a galaxy in a perfectly circular orbit, and the mean rotation speed of the stellar population under study $\overline{v_\phi}$. The asymmetric drift gives information about the tendency of a stellar population to lag behind other populations. The main reason for this lag can be found in the collisionless nature of the stars within a galaxy, which makes them sensitive to dynamical heating. In this kind of process the stars tend to migrate to orbits with larger radii, which leads to a decrease of the rotational velocity, as expected from angular momentum conservation. These changes are translated into an increase of the random motions of the stars and consequently an increase of the velocity dispersion of the stellar populations \citepads{2002MNRAS.336..785S}. The more time passes, the more heating events will take place, thus it is expected to find higher velocity dispersions in old stellar populations \citepads{1996ApJ...460..121W}.

\citetads{2016MNRAS.458.1199E} studied the slope of the inner rotation curves of a sample of spiral galaxies from the S$^4$G survey, using H$\alpha$ Fabry-Perot data to derive the rotation velocity of the ionised-gas component. They corrected these velocities for the asymmetric drift effect to obtain the circular velocity. While their results were estimated for ionised gas, they can nevertheless provide us with a reasonable approximation of the values we could expect for the stars. 

When we evaluate the results of \citetads{2016MNRAS.458.1199E} at a radius of $650$\,pc, which is approximately the average radius of the rings of the galaxies in our sample, we obtain an asymmetric drift of around $2$\,\kms. This value is very close to the average velocity differences estimated in the rings of our galaxies, but significantly smaller than the most extreme velocity differences estimated from our data. The asymmetric drift may thus have some relevance in the ring kinematics, especially concerning the velocity dispersion distribution, but it is not the main driver in those regions where we find the higher kinematic differences between the young and old stellar populations.

\subsection{The effect of Paschen lines}
\label{sec:paschen}

Another possible effect that may be altering the kinematics estimated from the MUSE data is the presence of the Paschen line series in absorption in the  Ca\,II Triplet region. The Paschen series fall in the near-infrared spectral range and some of the transitions overlap with the Ca\,II lines. We could expect to observe some dependence of this effect with age since the Paschen lines Pa\,16, Pa\,15 and Pa\,13 are strong in young stars. This superposition could result in a change in the absorption lines shape, which will tend to increase the measured velocity dispersion \citep[see, e.g., Fig.~10 in][]{2003MNRAS.340.1317V}. Given that \deltaS\ in the ring-dominated regions is such that velocity dispersions are larger in the H$\beta$ region than around the Ca\,II Triplet, it is unlikely that the effect of the Paschen lines plays any significant role in our measurements. 

\section{Summary and conclusions}
\label{sec:conclusions}

The goal of the present work is to explore the relation between the kinematics derived from the blue and red spectral ranges and the properties linked to different stellar populations harboured in the inner regions of four nearby galaxies in the TIMER sample. We derived the velocity and velocity dispersion around the H$\beta$ and the Ca\,{\sc II} Triplet spectral lines in these regions using the \textsc{GIST} pipeline. Additionally, we estimated the colour excess E(B$-$V). Assuming that the H$\beta$ line is sensitive to the young stellar population and that the Ca\,{\sc II} Triplet is dominated by the old stellar population, we analysed the correlation of the difference of velocity and velocity dispersion between both populations and the various other derived parameters. Our results can be summarised as follows:

\begin{itemize}
    \item The young stellar population presents not only a higher velocity, but also a higher velocity dispersion than the old stellar population. This is especially remarkable in the nuclear rings, where differences between both populations reach up to $70\pm7\,{\rm km\, s^{-1}}$ in velocity and $88\pm9\,{\rm km\, s^{-1}}$ velocity dispersion in the case of NGC\,1097. Nevertheless, the mean values of the difference in velocity in the rings are low (below $10\,{\rm km\, s^{-1}}$), while the mean differences in velocity dispersion are more significant, reaching up to $24\pm9\,{\rm km\, s^{-1}}$ in the most extreme case.

    \item We observe a clear trend of the difference of velocity dispersion increasing with extinction in the ring of NGC\,1097. This trend is also observed in the other rings, but is weaker there.
    
    \item We demonstrate with the ring of NGC\,1097 that the physical size of the bins plays a limited role in the change of the trend of \deltaS\ with colour excess. Our study suggests that beam-smearing effects could be the reason for weaker trends in the rings of the other galaxies in our sample.

    \item From a simulation exercise in which we mix a set of mock spectra of young and old stellar populations, we demonstrate that masking the H$\beta$ line produces a lower sensitivity when determining differences in velocity and velocity dispersion. This explains the lack of trends with age or fraction of young stars.
    
    \item We can rule out asymmetric drift or overlapping Paschen lines on top of the Ca\,{\sc II} ones as drivers of the observed kinematic differences.
\end{itemize}

Our analysis indicates that the kinematic differences observed are not strongly related to the presence of different populations. However, the presence of dust and the physical size of the bins in our maps play important roles in the difference of velocity dispersion. We demonstrate that, in extreme situations, these differences are observable at the intermediate resolution offered by the current generation of IFUs, such as MUSE. In the near future, with the next generation of high-spectral resolution spectrographs \citep[e.g. WEAVE;][]{weave}, it will be possible to perform even more detailed chemo-kinematic decompositions of nearby galaxies.

\begin{acknowledgements}
Based on observations collected at the European Southern Observatory under ESO programmes 097.B-0640(A). We acknowledge financial support from the European Union's Horizon 2020 research and innovation programme under Marie Sk\l odowska-Curie grant agreement No 721463 to the SUNDIAL ITN network, from the State Research Agency (AEI) of the Spanish Ministry of Science and Innovation and the European Regional Development Fund (FEDER) under the grant with reference PID2019-105602GB-I00, and from IAC project P/300724, financed by the Ministry of Science and Innovation, through the State Budget and by the Canary Islands Department of Economy, Knowledge and Employment, through the Regional Budget of the Autonomous Community. JF-B acknowledges support from grants AYA2016-77237-C3-1-P and PID2019-107427GB-C32 of the Spanish Ministry of Science, Innovation and Universities (MCIU) and through the IAC project TRACES, which is partially supported by the state budget and the regional budget of the Consejer\'ia de Econom\'ia, Industria, Comercio y Conocimiento of the Canary Islands Autonomous Community. The Science, Technology and Facilities Council is acknowledged by JN for support through the Consolidated Grant Cosmology and Astrophysics at Portsmouth, ST/S000550/1. JMA acknowledge support from the MCIU by the grant AYA2017-83204-P and the Programa Operativo FEDER Andaluc\'{i}a 2014-2020 in collaboration with the Andalucian Office for Economy and Knowledge. PC acknowledges financial support from Funda\c{c}\~{a}o de Amparo \`{a} Pesquisa do Estado de S\~{a}o Paulo (FAPESP) process number 2018/05392-8 and Conselho Nacional de Desenvolvimento Cient\'ifico e Tecnol\'ogico (CNPq) process number  310041/2018-0. GvdV acknowledges funding from the European Research Council (ERC) under the European Union's Horizon 2020 research and innovation programme under grant agreement No 724857 (Consolidator Grant ArcheoDyn). TK was supported by the Basic Science Research Program through the National Research Foundation of Korea (NRF) funded by the Ministry of Education(No. 2019R1A6A3A01092024).
\end{acknowledgements}


\bibliographystyle{aa}
\bibliography{biblio_rings}

\begin{thebibliography}{91}
\expandafter\ifx\csname natexlab\endcsname\relax\def\natexlab#1{#1}\fi

\bibitem[{{Aguerri} {et~al.}(2000){Aguerri}, {Mu{\~n}oz-Tu{\~n}{\'o}n},
  {Varela}, \& {Prieto}}]{2000A&A...361..841A}
{Aguerri}, J.~A.~L., {Mu{\~n}oz-Tu{\~n}{\'o}n}, C., {Varela}, A.~M., \&
  {Prieto}, M. 2000, \aap, 361, 841

\bibitem[{{Bacon} {et~al.}(2010){Bacon}, {Accardo}, {Adjali}, {Anwand},
  {Bauer}, {Biswas}, {Blaizot}, {Boudon}, {Brau-Nogue}, {Brinchmann},
  {Caillier}, {Capoani}, {Carollo}, {Contini}, {Couderc}, {Daguis{\'e}},
  {Deiries}, {Delabre}, {Dreizler}, {Dubois}, {Dupieux}, {Dupuy}, {Emsellem},
  {Fechner}, {Fleischmann}, {Fran{\c{c}}ois}, {Gallou}, {Gharsa}, {Glindemann},
  {Gojak}, {Guiderdoni}, {Hansali}, {Hahn}, {Jarno}, {Kelz}, {Koehler},
  {Kosmalski}, {Laurent}, {Le Floch}, {Lilly}, {Lizon}, {Loupias}, {Manescau},
  {Monstein}, {Nicklas}, {Olaya}, {Pares}, {Pasquini}, {P{\'e}contal-Rousset},
  {Pell{\'o}}, {Petit}, {Popow}, {Reiss}, {Remillieux}, {Renault}, {Roth},
  {Rupprecht}, {Serre}, {Schaye}, {Soucail}, {Steinmetz}, {Streicher}, {Stuik},
  {Valentin}, {Vernet}, {Weilbacher}, {Wisotzki}, \&
  {Yerle}}]{2010SPIE.7735E..08B}
{Bacon}, R., {Accardo}, M., {Adjali}, L., {et~al.} 2010, Society of
  Photo-Optical Instrumentation Engineers (SPIE) Conference Series, Vol. 7735,
  {The MUSE second-generation VLT instrument}, 773508

\bibitem[{{Bacon} {et~al.}(2017){Bacon}, {Conseil}, {Mary}, {Brinchmann},
  {Shepherd}, {Akhlaghi}, {Weilbacher}, {Piqueras}, {Wisotzki}, {Lagattuta},
  {Epinat}, {Guerou}, {Inami}, {Cantalupo}, {Courbot}, {Contini}, {Richard},
  {Maseda}, {Bouwens}, {Bouch{\'e}}, {Kollatschny}, {Schaye}, {Marino},
  {Pello}, {Herenz}, {Guiderdoni}, \& {Carollo}}]{2017A&A...608A...1B}
{Bacon}, R., {Conseil}, S., {Mary}, D., {et~al.} 2017, \aap, 608, A1

\bibitem[{{Bacon} {et~al.}(2001){Bacon}, {Copin}, {Monnet}, {Miller},
  {Allington-Smith}, {Bureau}, {Carollo}, {Davies}, {Emsellem}, {Kuntschner},
  {Peletier}, {Verolme}, \& {de Zeeuw}}]{2001MNRAS.326...23B}
{Bacon}, R., {Copin}, Y., {Monnet}, G., {et~al.} 2001, \mnras, 326, 23

\bibitem[{{Baes} {et~al.}(2003){Baes}, {Davies}, {Dejonghe}, {Sabatini},
  {Roberts}, {Evans}, {Linder}, {Smith}, \& {de Blok}}]{2003MNRAS.343.1081B}
{Baes}, M., {Davies}, J.~I., {Dejonghe}, H., {et~al.} 2003, \mnras, 343, 1081

\bibitem[{{Baes} \& {Dejonghe}(2000)}]{2000MNRAS.313..153B}
{Baes}, M. \& {Dejonghe}, H. 2000, \mnras, 313, 153

\bibitem[{{Baes} {et~al.}(2000){Baes}, {Dejonghe}, \& {De
  Rijcke}}]{2000MNRAS.318..798B}
{Baes}, M., {Dejonghe}, H., \& {De Rijcke}, S. 2000, \mnras, 318, 798

\bibitem[{{Binney} \& {Merrifield}(1998)}]{1998gaas.book.....B}
{Binney}, J. \& {Merrifield}, M. 1998, {Galactic Astronomy}

\bibitem[{{Bittner} {et~al.}(2019){Bittner}, {Falc{\'o}n-Barroso}, {Nedelchev},
  {Dorta}, {Gadotti}, {Sarzi}, {Molaeinezhad}, {Iodice}, {Rosado-Belza}, {de
  Lorenzo-C{\'a}ceres}, {Fragkoudi}, {Gal{\'a}n-de Anta}, {Husemann},
  {M{\'e}ndez-Abreu}, {Neumann}, {Pinna}, {Querejeta},
  {S{\'a}nchez-Bl{\'a}zquez}, \& {Seidel}}]{2019A&A...628A.117B}
{Bittner}, A., {Falc{\'o}n-Barroso}, J., {Nedelchev}, B., {et~al.} 2019, \aap,
  628, A117

\bibitem[{{Bittner} {et~al.}(2020){Bittner}, {S{\'a}nchez-Bl{\'a}zquez},
  {Gadotti}, {Neumann}, {Fragkoudi}, {Coelho}, {de Lorenzo-C{\'a}ceres},
  {Falc{\'o}n-Barroso}, {Kim}, {Leaman}, {Mart{\'\i}n-Navarro},
  {M{\'e}ndez-Abreu}, {P{\'e}rez}, {Querejeta}, {Seidel}, \& {van de
  Ven}}]{2020arXiv200901856B}
{Bittner}, A., {S{\'a}nchez-Bl{\'a}zquez}, P., {Gadotti}, D.~A., {et~al.} 2020,
  arXiv e-prints, arXiv:2009.01856

\bibitem[{{B{\"o}ker} {et~al.}(2008){B{\"o}ker}, {Falc{\'o}n-Barroso},
  {Schinnerer}, {Knapen}, \& {Ryder}}]{2008AJ....135..479B}
{B{\"o}ker}, T., {Falc{\'o}n-Barroso}, J., {Schinnerer}, E., {Knapen}, J.~H.,
  \& {Ryder}, S. 2008, \aj, 135, 479

\bibitem[{{Bonnet} {et~al.}(2004){Bonnet}, {Abuter}, {Baker}, {Bornemann},
  {Brown}, {Castillo}, {Conzelmann}, {Damster}, {Davies}, {Delabre},
  {Donaldson}, {Dumas}, {Eisenhauer}, {Elswijk}, {Fedrigo}, {Finger},
  {Gemperlein}, {Genzel}, {Gilbert}, {Gillet}, {Goldbrunner}, {Horrobin}, {Ter
  Horst}, {Huber}, {Hubin}, {Iserlohe}, {Kaufer}, {Kissler-Patig}, {Kragt},
  {Kroes}, {Lehnert}, {Lieb}, {Liske}, {Lizon}, {Lutz}, {Modigliani}, {Monnet},
  {Nesvadba}, {Patig}, {Pragt}, {Reunanen}, {R{\"o}hrle}, {Rossi}, {Schmutzer},
  {Schoenmaker}, {Schreiber}, {Stroebele}, {Szeifert}, {Tacconi}, {Tecza},
  {Thatte}, {Tordo}, {van der Werf}, \& {Weisz}}]{2004Msngr.117...17B}
{Bonnet}, H., {Abuter}, R., {Baker}, A., {et~al.} 2004, The Messenger, 117, 17

\bibitem[{{Bosma} {et~al.}(1992){Bosma}, {Byun}, {Freeman}, \&
  {Athanassoula}}]{1992ApJ...400L..21B}
{Bosma}, A., {Byun}, Y., {Freeman}, K.~C., \& {Athanassoula}, E. 1992, \apjl,
  400, L21

\bibitem[{{Bruzual} \& {Charlot}(2003)}]{2003MNRAS.344.1000B}
{Bruzual}, G. \& {Charlot}, S. 2003, \mnras, 344, 1000

\bibitem[{{Bundy} {et~al.}(2015){Bundy}, {Bershady}, {Law}, {Yan}, {Drory},
  {MacDonald}, {Wake}, {Cherinka}, {S{\'a}nchez-Gallego}, {Weijmans}, {Thomas},
  {Tremonti}, {Masters}, {Coccato}, {Diamond-Stanic}, {Arag{\'o}n-Salamanca},
  {Avila-Reese}, {Badenes}, {Falc{\'o}n-Barroso}, {Belfiore}, {Bizyaev},
  {Blanc}, {Bland-Hawthorn}, {Blanton}, {Brownstein}, {Byler}, {Cappellari},
  {Conroy}, {Dutton}, {Emsellem}, {Etherington}, {Frinchaboy}, {Fu}, {Gunn},
  {Harding}, {Johnston}, {Kauffmann}, {Kinemuchi}, {Klaene}, {Knapen},
  {Leauthaud}, {Li}, {Lin}, {Maiolino}, {Malanushenko}, {Malanushenko}, {Mao},
  {Maraston}, {McDermid}, {Merrifield}, {Nichol}, {Oravetz}, {Pan}, {Parejko},
  {Sanchez}, {Schlegel}, {Simmons}, {Steele}, {Steinmetz}, {Thanjavur},
  {Thompson}, {Tinker}, {van den Bosch}, {Westfall}, {Wilkinson}, {Wright},
  {Xiao}, \& {Zhang}}]{2015ApJ...798....7B}
{Bundy}, K., {Bershady}, M.~A., {Law}, D.~R., {et~al.} 2015, \apj, 798, 7

\bibitem[{{Buta} \& {Combes}(1996)}]{1996FCPh...17...95B}
{Buta}, R. \& {Combes}, F. 1996, \fcp, 17, 95

\bibitem[{{Buta} {et~al.}(2015){Buta}, {Sheth}, {Athanassoula}, {Bosma},
  {Knapen}, {Laurikainen}, {Salo}, {Elmegreen}, {Ho}, {Zaritsky}, {Courtois},
  {Hinz}, {Mu{\~n}oz-Mateos}, {Kim}, {Regan}, {Gadotti}, {Gil de Paz}, {Laine},
  {Men{\'e}ndez-Delmestre}, {Comer{\'o}n}, {Erroz Ferrer}, {Seibert},
  {Mizusawa}, {Holwerda}, \& {Madore}}]{2015ApJS..217...32B}
{Buta}, R.~J., {Sheth}, K., {Athanassoula}, E., {et~al.} 2015, \apjs, 217, 32

\bibitem[{{Cappellari} {et~al.}(2011){Cappellari}, {Emsellem}, {Krajnovi{\'c}},
  {McDermid}, {Scott}, {Verdoes Kleijn}, {Young}, {Alatalo}, {Bacon}, {Blitz},
  {Bois}, {Bournaud}, {Bureau}, {Davies}, {Davis}, {de Zeeuw}, {Duc},
  {Khochfar}, {Kuntschner}, {Lablanche}, {Morganti}, {Naab}, {Oosterloo},
  {Sarzi}, {Serra}, \& {Weijmans}}]{2011MNRAS.413..813C}
{Cappellari}, M., {Emsellem}, E., {Krajnovi{\'c}}, D., {et~al.} 2011, \mnras,
  413, 813

\bibitem[{{Carrasco} {et~al.}(2018){Carrasco}, {Gil de Paz}, {Gallego},
  {Iglesias-P{\'a}ramo}, {Cedazo}, {Garc{\'\i}a Vargas}, {Arrillaga},
  {Avil{\'e}s}, {Bouquin}, {Carbajo}, {Cardiel}, {Carrera}, {Castillo Morales},
  {Castillo-Dom{\'\i}nguez}, {Esteban San Rom{\'a}n}, {Ferrusca},
  {G{\'o}mez-{\'A}lvarez}, {Izazaga-P{\'e}rez}, {Lefort}, {L{\'o}pez Orozco},
  {Maldonado}, {Mart{\'\i}nez Delgado}, {Morales Dur{\'a}n}, {M{\'u}jica},
  {Ortiz}, {P{\'a}ez}, {Pascual}, {P{\'e}rez-Calpena}, {Picazo},
  {S{\'a}nchez-Penim}, {S{\'a}nchez-Blanco}, {Tulloch}, {Vel{\'a}zquez},
  {V{\'\i}lchez}, {Zamorano}, {Aguerri}, {Barrado}, {Bertone}, {Cava},
  {Catal{\'a}n-Torrecilla}, {Cenarro}, {Ch{\'a}vez}, {Dullo}, {Eliche},
  {Garc{\'\i}a}, {Garc{\'\i}a-Rojas}, {Guichard}, {Gonz{\'a}lez-Delgado},
  {Guzm{\'a}n}, {Herrero}, {Hu{\'e}lamo}, {Hughes}, {Jim{\'e}nez-Vicente},
  {Kehrig}, {Marino}, {M{\'a}rquez}, {Masegosa}, {Mayya}, {M{\'e}ndez-Abreu},
  {Moll{\'a}}, {Mu{\~n}oz-Tu{\~n}{\'o}n}, {Peimbert}, {P{\'e}rez-Gonz{\'a}lez},
  {P{\'e}rez-Montero}, {Roca-F{\`a}brega}, {Rodr{\'\i}guez},
  {Rodr{\'\i}guez-Espinosa}, {Rodr{\'\i}guez-Merino},
  {Rodr{\'\i}guez-Mu{\~n}oz}, {Rosa-Gonz{\'a}lez}, {S{\'a}nchez-Almeida},
  {S{\'a}nchez Contreras}, {S{\'a}nchez-Bl{\'a}zquez}, {S{\'a}nchez},
  {Sarajedini}, {Silich}, {Sim{\'o}n-D{\'\i}az}, {Tenorio-Tagle}, {Terlevich},
  {Terlevich}, {Torres-Peimbert}, {Trujillo}, {Tsamis}, \&
  {Vega}}]{2018SPIE10702E..16C}
{Carrasco}, E., {Gil de Paz}, A., {Gallego}, J., {et~al.} 2018, in Society of
  Photo-Optical Instrumentation Engineers (SPIE) Conference Series, Vol. 10702,
  \procspie, 1070216

\bibitem[{{Chilingarian} {et~al.}(2007){Chilingarian}, {Prugniel},
  {Sil'Chenko}, \& {Koleva}}]{2007IAUS..241..175C}
{Chilingarian}, I., {Prugniel}, P., {Sil'Chenko}, O., \& {Koleva}, M. 2007, in
  IAU Symposium, Vol. 241, Stellar Populations as Building Blocks of Galaxies,
  ed. A.~{Vazdekis} \& R.~{Peletier}, 175--176

\bibitem[{Cleveland \& Devlin(1988)}]{Clev:Devl:1988}
Cleveland, W.~S. \& Devlin, S.~J. 1988, Journal of the American Statistical
  Association, 83, 596

\bibitem[{{Coccato} {et~al.}(2013){Coccato}, {Morelli}, {Pizzella}, {Corsini},
  {Buson}, \& {Dalla Bont{\`a}}}]{2013A&A...549A...3C}
{Coccato}, L., {Morelli}, L., {Pizzella}, A., {et~al.} 2013, \aap, 549, A3

\bibitem[{{Combes} \& {Gerin}(1985)}]{1985A&A...150..327C}
{Combes}, F. \& {Gerin}, M. 1985, \aap, 150, 327

\bibitem[{{Combes} \& {Sanders}(1981)}]{1981A&A....96..164C}
{Combes}, F. \& {Sanders}, R.~H. 1981, \aap, 96, 164

\bibitem[{{Comer{\'o}n} {et~al.}(2010){Comer{\'o}n}, {Knapen}, {Beckman},
  {Laurikainen}, {Salo}, {Mart{\'\i}nez-Valpuesta}, \&
  {Buta}}]{2010MNRAS.402.2462C}
{Comer{\'o}n}, S., {Knapen}, J.~H., {Beckman}, J.~E., {et~al.} 2010, \mnras,
  402, 2462

\bibitem[{{Croom} {et~al.}(2012){Croom}, {Lawrence}, {Bland-Hawthorn},
  {Bryant}, {Fogarty}, {Richards}, {Goodwin}, {Farrell}, {Miziarski}, {Heald},
  {Jones}, {Lee}, {Colless}, {Brough}, {Hopkins}, {Bauer}, {Birchall}, {Ellis},
  {Horton}, {Leon-Saval}, {Lewis}, {L{\'o}pez-S{\'a}nchez}, {Min}, {Trinh}, \&
  {Trowland}}]{2012MNRAS.421..872C}
{Croom}, S.~M., {Lawrence}, J.~S., {Bland-Hawthorn}, J., {et~al.} 2012, \mnras,
  421, 872

\bibitem[{{Dalton} {et~al.}(2018{\natexlab{a}}){Dalton}, {Trager}, {Abrams},
  {Bonifacio}, {Aguerri}, {Vallenari}, {Middleton}, {Benn}, {Dee},
  {Say{\`e}de}, {Lewis}, {Pragt}, {Pic{\'o}}, {Walton}, {Rey}, {Allende},
  {Lhom{\'e}}, {Terrett}, {Brock}, {Gilbert}, {Ridings}, {Verheijen}, {Tosh},
  {Steele}, {Stuik}, {Kroes}, {Tromp}, {Kragt}, {Lesman}, {Mottram}, {Bates},
  {Gribbin}, {Burgal}, {Herreros}, {Delgado}, {Martin}, {Cano}, {Navarro},
  {Irwin}, {Lewis}, {Gonzales Solares}, {O'Mahony}, {Bianco}, {Zurita}, {ter
  Horst}, {Molinari}, {Lodi}, {Guerra}, {Baruffolo}, {Carrasco}, {Farkas},
  {Schallig}, {Hill}, {Smith}, {Drew}, {Poggianti}, {Pieri}, {Jin}, {Dominquez
  Palmero}, {Fari{\~n}a}, {Martin}, {Worley}, {Murphy}, {Hidalgo}, {Mignot},
  {Bishop}, {Guest}, {Elswijk}, {de Haan}, {Hanenburg}, {Salasnich}, {Mayya},
  {Izazaga-P{\'e}rez}, \& {Peralta de Arriba}}]{weave}
{Dalton}, G., {Trager}, S., {Abrams}, D.~C., {et~al.} 2018{\natexlab{a}}, in
  Society of Photo-Optical Instrumentation Engineers (SPIE) Conference Series,
  Vol. 10702, Ground-based and Airborne Instrumentation for Astronomy VII,
  107021B

\bibitem[{{Dalton} {et~al.}(2018{\natexlab{b}}){Dalton}, {Trager}, {Abrams},
  {Bonifacio}, {Aguerri}, {Vallenari}, {Middleton}, {Benn}, {Dee},
  {Say{\`e}de}, {Lewis}, {Pragt}, {Pic{\'o}}, {Walton}, {Rey}, {Allende},
  {Lhom{\'e}}, {Terrett}, {Brock}, {Gilbert}, {Ridings}, {Verheijen}, {Tosh},
  {Steele}, {Stuik}, {Kroes}, {Tromp}, {Kragt}, {Lesman}, {Mottram}, {Bates},
  {Gribbin}, {Burgal}, {Herreros}, {Delgado}, {Martin}, {Cano}, {Navarro},
  {Irwin}, {Lewis}, {Gonzales Solares}, {O'Mahony}, {Bianco}, {Zurita}, {ter
  Horst}, {Molinari}, {Lodi}, {Guerra}, {Baruffolo}, {Carrasco}, {Farkas},
  {Schallig}, {Hill}, {Smith}, {Drew}, {Poggianti}, {Pieri}, {Jin}, {Dominquez
  Palmero}, {Fari{\~n}a}, {Martin}, {Worley}, {Murphy}, {Hidalgo}, {Mignot},
  {Bishop}, {Guest}, {Elswijk}, {de Haan}, {Hanenburg}, {Salasnich}, {Mayya},
  {Izazaga-P{\'e}rez}, \& {Peralta de Arriba}}]{2018SPIE10702E..1BD}
{Dalton}, G., {Trager}, S., {Abrams}, D.~C., {et~al.} 2018{\natexlab{b}}, in
  Society of Photo-Optical Instrumentation Engineers (SPIE) Conference Series,
  Vol. 10702, Ground-based and Airborne Instrumentation for Astronomy VII,
  107021B

\bibitem[{{Davies}(1990)}]{1990MNRAS.245..350D}
{Davies}, J.~I. 1990, \mnras, 245, 350

\bibitem[{{Eisenhauer} {et~al.}(2003){Eisenhauer}, {Abuter}, {Bickert},
  {Biancat-Marchet}, {Bonnet}, {Brynnel}, {Conzelmann}, {Delabre}, {Donaldson},
  {Farinato}, {Fedrigo}, {Genzel}, {Hubin}, {Iserlohe}, {Kasper},
  {Kissler-Patig}, {Monnet}, {Roehrle}, {Schreiber}, {Stroebele}, {Tecza},
  {Thatte}, \& {Weisz}}]{2003SPIE.4841.1548E}
{Eisenhauer}, F., {Abuter}, R., {Bickert}, K., {et~al.} 2003, Society of
  Photo-Optical Instrumentation Engineers (SPIE) Conference Series, Vol. 4841,
  {SINFONI - Integral field spectroscopy at 50 milli-arcsecond resolution with
  the ESO VLT}, ed. M.~{Iye} \& A.~F.~M. {Moorwood}, 1548--1561

\bibitem[{{Erroz-Ferrer} {et~al.}(2016){Erroz-Ferrer}, {Knapen}, {Leaman},
  {D{\'\i}az-Garc{\'\i}a}, {Salo}, {Laurikainen}, {Querejeta},
  {Mu{\~n}oz-Mateos}, {Athanassoula}, {Bosma}, {Comer{\'o}n}, {Elmegreen}, \&
  {Mart{\'\i}nez-Valpuesta}}]{2016MNRAS.458.1199E}
{Erroz-Ferrer}, S., {Knapen}, J.~H., {Leaman}, R., {et~al.} 2016, \mnras, 458,
  1199

\bibitem[{{Fabricius} {et~al.}(2008){Fabricius}, {Barnes}, {Bender}, {Drory},
  {Grupp}, {Hill}, {Hopp}, \& {MacQueen}}]{2008SPIE.7014E..73F}
{Fabricius}, M.~H., {Barnes}, S., {Bender}, R., {et~al.} 2008, in Society of
  Photo-Optical Instrumentation Engineers (SPIE) Conference Series, Vol. 7014,
  Ground-based and Airborne Instrumentation for Astronomy II, 701473

\bibitem[{{Falc{\'o}n-Barroso} {et~al.}(2006){Falc{\'o}n-Barroso}, {Bacon},
  {Bureau}, {Cappellari}, {Davies}, {de Zeeuw}, {Emsellem}, {Fathi},
  {Krajnovi{\'c}}, {Kuntschner}, {McDermid}, {Peletier}, \&
  {Sarzi}}]{2006MNRAS.369..529F}
{Falc{\'o}n-Barroso}, J., {Bacon}, R., {Bureau}, M., {et~al.} 2006, \mnras,
  369, 529

\bibitem[{{Falc{\'o}n-Barroso} {et~al.}(2011){Falc{\'o}n-Barroso},
  {S{\'a}nchez-Bl{\'a}zquez}, {Vazdekis}, {Ricciardelli}, {Cardiel}, {Cenarro},
  {Gorgas}, \& {Peletier}}]{2011A&A...532A..95F}
{Falc{\'o}n-Barroso}, J., {S{\'a}nchez-Bl{\'a}zquez}, P., {Vazdekis}, A.,
  {et~al.} 2011, \aap, 532, A95

\bibitem[{{Gadotti} {et~al.}(2019){Gadotti}, {S{\'a}nchez-Bl{\'a}zquez},
  {Falc{\'o}n-Barroso}, {Husemann}, {Seidel}, {P{\'e}rez}, {de
  Lorenzo-C{\'a}ceres}, {Martinez-Valpuesta}, {Fragkoudi}, {Leung}, {van de
  Ven}, {Leaman}, {Coelho}, {Martig}, {Kim}, {Neumann}, \&
  {Querejeta}}]{2019MNRAS.482..506G}
{Gadotti}, D.~A., {S{\'a}nchez-Bl{\'a}zquez}, P., {Falc{\'o}n-Barroso}, J.,
  {et~al.} 2019, \mnras, 482, 506

\bibitem[{{Gadotti} {et~al.}(2015){Gadotti}, {Seidel},
  {S{\'a}nchez-Bl{\'a}zquez}, {Falc{\'o}n-Barroso}, {Husemann}, {Coelho}, \&
  {P{\'e}rez}}]{2015A&A...584A..90G}
{Gadotti}, D.~A., {Seidel}, M.~K., {S{\'a}nchez-Bl{\'a}zquez}, P., {et~al.}
  2015, \aap, 584, A90

\bibitem[{{Gil de Paz} {et~al.}(2018){Gil de Paz}, {Carrasco}, {Gallego},
  {Iglesias-P{\'a}ramo}, {Cedazo}, {Garc{\'\i}a-Vargas}, {Arrillaga},
  {Avil{\'e}s}, {Bouquin}, {Carbajo}, {Cardiel}, {Carrera}, {Castillo-Morales},
  {Castillo-Dom{\'\i}nguez}, {Esteban San Rom{\'a}n}, {Ferrusca},
  {G{\'o}mez-{\'A}lvarez}, {Izazaga-P{\'e}rez}, {Lefort}, {L{\'o}pez-Orozco},
  {Maldonado}, {Mart{\'\i}nez-Delgado}, {Morales-Dur{\'a}n}, {Mujica},
  {P{\'a}ez}, {Pascual}, {P{\'e}rez-Calpena}, {Picazo}, {S{\'a}nchez-Penim},
  {S{\'a}nchez-Blanco}, {Tulloch}, {Vel{\'a}zquez}, {V{\'\i}lchez}, {Zamorano},
  {Aguerri}, {Barrado y Navascues}, {Berlanas}, {Bertone}, {Cava},
  {Catal{\'a}n-Torrecilla}, {Cenarro}, {Ch{\'a}vez}, {Dullo}, {Garc{\'\i}a},
  {Garc{\'\i}a-Rojas}, {Guichard}, {Gonz{\'a}lez-Delgado}, {Guzm{\'a}n},
  {Herrero}, {Hu{\'e}lamo}, {Hughes}, {Jim{\'e}nez-Vicente}, {Kehrig},
  {Marino}, {M{\'a}rquez}, {Masegosa}, {Mayya}, {M{\'e}ndez-Abreu},
  {Moll{\'a}}, {Mu{\~n}oz-Tu{\~n}{\'o}n}, {Peimbert}, {P{\'e}rez-Gonz{\'a}lez},
  {P{\'e}rez-Montero}, {Rodr{\'\i}guez}, {Rodr{\'\i}guez-Espinosa},
  {Rodr{\'\i}guez Merino}, {Rodr{\'\i}guez-Mu{\~n}oz}, {Rosa-Gonz{\'a}lez},
  {S{\'a}nchez-Almeida}, {S{\'a}nchez-Contreras}, {S{\'a}nchez-Bl{\'a}zquez},
  {S{\'a}nchez}, {Sarajedini}, {Silich}, {Sim{\'o}n-D{\'\i}az},
  {Tenorio-Tagle}, {Terlevich}, {Terlevich}, {Torres-Peimbert}, {Trujillo},
  {Tsamis}, \& {Vega}}]{2018SPIE10702E..17G}
{Gil de Paz}, A., {Carrasco}, E., {Gallego}, J., {et~al.} 2018, in Society of
  Photo-Optical Instrumentation Engineers (SPIE) Conference Series, Vol. 10702,
  \procspie, 1070217

\bibitem[{{Gil de Paz} {et~al.}(2014){Gil de Paz}, {Gallego}, {Carrasco},
  {Iglesias-P{\'a}ramo}, {Cedazo}, {V{\'\i}lchez}, {Garc{\'\i}a-Vargas},
  {Arrillaga}, {Carrera}, {Castillo-Morales}, {Castillo-Dom{\'\i}nguez},
  {Eliche-Moral}, {Ferrusca}, {Gonz{\'a}lez-Guardia}, {Lefort}, {Maldonado},
  {Marino}, {Mart{\'\i}nez-Delgado}, {Morales Dur{\'a}n}, {Mujica}, {P{\'a}ez},
  {Pascual}, {P{\'e}rez-Calpena}, {S{\'a}nchez-Penim}, {S{\'a}nchez-Blanco},
  {Tulloch}, {Vel{\'a}zquez}, {Zamorano}, {Aguerri}, {Barrado y Nav{\'a}scues},
  {Bertone}, {Cardiel}, {Cava}, {Cenarro}, {Ch{\'a}vez}, {Garc{\'\i}a},
  {Guichard}, {G{\'u}zman}, {Herrero}, {Hu{\'e}lamo}, {Hughes},
  {Jim{\'e}nez-Vicente}, {Kehrig}, {M{\'a}rquez}, {Masegosa}, {Mayya},
  {M{\'e}ndez-Abreu}, {Moll{\'a}}, {Mu{\~n}oz-Tu{\~n}{\'o}n}, {Peimbert},
  {P{\'e}rez-Gonz{\'a}lez}, {P{\'e}rez Montero}, {Rodr{\'\i}guez},
  {Rodr{\'\i}guez-Espinosa}, {Rodr{\'\i}guez-Merino}, {Rosa-Gonz{\'a}lez},
  {S{\'a}nchez-Almeida}, {S{\'a}nchez Contreras}, {S{\'a}nchez-Bl{\'a}zquez},
  {S{\'a}nchez Moreno}, {S{\'a}nchez}, {Sarajedini}, {Serena}, {Silich},
  {Sim{\'o}n-D{\'\i}az}, {Tenorio-Tagle}, {Terlevich}, {Terlevich},
  {Torres-Peimbert}, {Trujillo}, {Tsamis}, {Vega}, \&
  {Villar}}]{2014SPIE.9147E..0OG}
{Gil de Paz}, A., {Gallego}, J., {Carrasco}, E., {et~al.} 2014, in Society of
  Photo-Optical Instrumentation Engineers (SPIE) Conference Series, Vol. 9147,
  \procspie, 91470O

\bibitem[{{Gu{\'e}rou} {et~al.}(2017){Gu{\'e}rou}, {Krajnovi{\'c}}, {Epinat},
  {Contini}, {Emsellem}, {Bouch{\'e}}, {Bacon}, {Michel-Dansac}, {Richard},
  {Weilbacher}, {Schaye}, {Marino}, {den Brok}, \&
  {Erroz-Ferrer}}]{2017A&A...608A...5G}
{Gu{\'e}rou}, A., {Krajnovi{\'c}}, D., {Epinat}, B., {et~al.} 2017, \aap, 608,
  A5

\bibitem[{{Johnston} {et~al.}(2018{\natexlab{a}}){Johnston}, {Hau}, {Coccato},
  \& {Herrera}}]{2018MNRAS.480.3215J}
{Johnston}, E.~J., {Hau}, G. K.~T., {Coccato}, L., \& {Herrera}, C.
  2018{\natexlab{a}}, \mnras, 480, 3215

\bibitem[{{Johnston} {et~al.}(2017){Johnston}, {H{\"a}u{\ss}ler},
  {Arag{\'o}n-Salamanca}, {Merrifield}, {Bamford}, {Bershady}, {Bundy},
  {Drory}, {Fu}, {Law}, {Nitschelm}, {Thomas}, {Roman Lopes}, {Wake}, \&
  {Yan}}]{2017MNRAS.465.2317J}
{Johnston}, E.~J., {H{\"a}u{\ss}ler}, B., {Arag{\'o}n-Salamanca}, A., {et~al.}
  2017, \mnras, 465, 2317

\bibitem[{{Johnston} {et~al.}(2018{\natexlab{b}}){Johnston}, {Merrifield}, \&
  {Arag{\'o}n-Salamanca}}]{2018MNRAS.478.4255J}
{Johnston}, E.~J., {Merrifield}, M., \& {Arag{\'o}n-Salamanca}, A.
  2018{\natexlab{b}}, \mnras, 478, 4255

\bibitem[{{Johnston} {et~al.}(2013){Johnston}, {Merrifield},
  {Arag{\'o}n-Salamanca}, \& {Cappellari}}]{2013MNRAS.428.1296J}
{Johnston}, E.~J., {Merrifield}, M.~R., {Arag{\'o}n-Salamanca}, A., \&
  {Cappellari}, M. 2013, \mnras, 428, 1296

\bibitem[{{Kelz} {et~al.}(2006){Kelz}, {Verheijen}, {Roth}, {Bauer}, {Becker},
  {Paschke}, {Popow}, {S{\'a}nchez}, \& {Laux}}]{2006PASP..118..129K}
{Kelz}, A., {Verheijen}, M. A.~W., {Roth}, M.~M., {et~al.} 2006, \pasp, 118,
  129

\bibitem[{{Knapen} {et~al.}(1995){Knapen}, {Beckman}, {Heller}, {Shlosman}, \&
  {de Jong}}]{1995ApJ...454..623K}
{Knapen}, J.~H., {Beckman}, J.~E., {Heller}, C.~H., {Shlosman}, I., \& {de
  Jong}, R.~S. 1995, \apj, 454, 623

\bibitem[{{Kormendy}(1979)}]{1979ApJ...227..714K}
{Kormendy}, J. 1979, \apj, 227, 714

\bibitem[{{Kormendy}(2013)}]{2013seg..book....1K}
{Kormendy}, J. 2013, {Secular Evolution in Disk Galaxies}, ed.
  J.~{Falc{\'o}n-Barroso} \& J.~H. {Knapen}, 1

\bibitem[{{Kormendy} \& {Kennicutt}(2004)}]{2004ARA&A..42..603K}
{Kormendy}, J. \& {Kennicutt}, Robert~C., J. 2004, \araa, 42, 603

\bibitem[{{Laine} {et~al.}(2001){Laine}, {Knapen}, {P{\'e}rez-Ram{\'\i}rez},
  {Englmaier}, \& {Matthias}}]{2001MNRAS.324..891L}
{Laine}, S., {Knapen}, J.~H., {P{\'e}rez-Ram{\'\i}rez}, D., {Englmaier}, P., \&
  {Matthias}, M. 2001, \mnras, 324, 891

\bibitem[{{Le F{\`e}vre} {et~al.}(2003){Le F{\`e}vre}, {Saisse}, {Mancini},
  {Brau-Nogue}, {Caputi}, {Castinel}, {D'Odorico}, {Garilli}, {Kissler-Patig},
  {Lucuix}, {Mancini}, {Pauget}, {Sciarretta}, {Scodeggio}, {Tresse}, \&
  {Vettolani}}]{2003SPIE.4841.1670L}
{Le F{\`e}vre}, O., {Saisse}, M., {Mancini}, D., {et~al.} 2003, in Society of
  Photo-Optical Instrumentation Engineers (SPIE) Conference Series, Vol. 4841,
  \procspie, ed. M.~{Iye} \& A.~F.~M. {Moorwood}, 1670--1681

\bibitem[{{Lindblad} \& {Kristen}(1996)}]{1996A&A...313..733L}
{Lindblad}, P.~A.~B. \& {Kristen}, H. 1996, \aap, 313, 733

\bibitem[{{Lindblad} {et~al.}(1997){Lindblad}, {Kristen}, {Joersaeter}, \&
  {Hoegbom}}]{1997A&A...317...36L}
{Lindblad}, P.~A.~B., {Kristen}, H., {Joersaeter}, S., \& {Hoegbom}, J. 1997,
  \aap, 317, 36

\bibitem[{{Maoz} {et~al.}(2001){Maoz}, {Barth}, {Ho}, {Sternberg}, \&
  {Filippenko}}]{2001AJ....121.3048M}
{Maoz}, D., {Barth}, A.~J., {Ho}, L.~C., {Sternberg}, A., \& {Filippenko},
  A.~V. 2001, \aj, 121, 3048

\bibitem[{{Maoz} {et~al.}(1996){Maoz}, {Barth}, {Sternberg}, {Filippenko},
  {Ho}, {Macchetto}, {Rix}, \& {Schneider}}]{1996AJ....111.2248M}
{Maoz}, D., {Barth}, A.~J., {Sternberg}, A., {et~al.} 1996, \aj, 111, 2248

\bibitem[{{Matthews} \& {Wood}(2001)}]{2001ApJ...548..150M}
{Matthews}, L.~D. \& {Wood}, K. 2001, \apj, 548, 150

\bibitem[{McDermid(2002)}]{etheses4179}
McDermid, R.~M. 2002, PhD thesis, Durham University

\bibitem[{{M{\'e}ndez-Abreu} {et~al.}(2019{\natexlab{a}}){M{\'e}ndez-Abreu},
  {S{\'a}nchez}, \& {de Lorenzo-C{\'a}ceres}}]{2019MNRAS.484.4298M}
{M{\'e}ndez-Abreu}, J., {S{\'a}nchez}, S.~F., \& {de Lorenzo-C{\'a}ceres}, A.
  2019{\natexlab{a}}, \mnras, 484, 4298

\bibitem[{{M{\'e}ndez-Abreu} {et~al.}(2019{\natexlab{b}}){M{\'e}ndez-Abreu},
  {S{\'a}nchez}, \& {de Lorenzo-C{\'a}ceres}}]{2019MNRAS.488L..80M}
{M{\'e}ndez-Abreu}, J., {S{\'a}nchez}, S.~F., \& {de Lorenzo-C{\'a}ceres}, A.
  2019{\natexlab{b}}, \mnras, 488, L80

\bibitem[{{Mu{\~n}oz-Mateos} {et~al.}(2013){Mu{\~n}oz-Mateos}, {Sheth}, {Gil de
  Paz}, {Meidt}, {Athanassoula}, {Bosma}, {Comer{\'o}n}, {Elmegreen},
  {Elmegreen}, {Erroz-Ferrer}, {Gadotti}, {Hinz}, {Ho}, {Holwerda}, {Jarrett},
  {Kim}, {Knapen}, {Laine}, {Laurikainen}, {Madore}, {Menendez-Delmestre},
  {Mizusawa}, {Regan}, {Salo}, {Schinnerer}, {Seibert}, {Skibba}, \&
  {Zaritsky}}]{2013ApJ...771...59M}
{Mu{\~n}oz-Mateos}, J.~C., {Sheth}, K., {Gil de Paz}, A., {et~al.} 2013, \apj,
  771, 59

\bibitem[{{Mu{\~n}oz-Mateos} {et~al.}(2015){Mu{\~n}oz-Mateos}, {Sheth},
  {Regan}, {Kim}, {Laine}, {Erroz-Ferrer}, {Gil de Paz}, {Comeron}, {Hinz},
  {Laurikainen}, {Salo}, {Athanassoula}, {Bosma}, {Bouquin}, {Schinnerer},
  {Ho}, {Zaritsky}, {Gadotti}, {Madore}, {Holwerda}, {Men{\'e}ndez-Delmestre},
  {Knapen}, {Meidt}, {Querejeta}, {Mizusawa}, {Seibert}, {Laine}, \&
  {Courtois}}]{2015ApJS..219....3M}
{Mu{\~n}oz-Mateos}, J.~C., {Sheth}, K., {Regan}, M., {et~al.} 2015, \apjs, 219,
  3

\bibitem[{{Ocvirk} {et~al.}(2006){Ocvirk}, {Pichon}, {Lan{\c{c}}on}, \&
  {Thi{\'e}baut}}]{2006MNRAS.365...74O}
{Ocvirk}, P., {Pichon}, C., {Lan{\c{c}}on}, A., \& {Thi{\'e}baut}, E. 2006,
  \mnras, 365, 74

\bibitem[{{Paturel} {et~al.}(2003){Paturel}, {Petit}, {Prugniel}, {Theureau},
  {Rousseau}, {Brouty}, {Dubois}, \& {Cambr{\'e}sy}}]{2003A&A...412...45P}
{Paturel}, G., {Petit}, C., {Prugniel}, P., {et~al.} 2003, \aap, 412, 45

\bibitem[{{P{\'e}rez-Ram{\'\i}rez} {et~al.}(2000){P{\'e}rez-Ram{\'\i}rez},
  {Knapen}, {Peletier}, {Laine}, {Doyon}, \& {Nadeau}}]{2000MNRAS.317..234P}
{P{\'e}rez-Ram{\'\i}rez}, D., {Knapen}, J.~H., {Peletier}, R.~F., {et~al.}
  2000, \mnras, 317, 234

\bibitem[{{Pillepich} {et~al.}(2018){Pillepich}, {Springel}, {Nelson}, {Genel},
  {Naiman}, {Pakmor}, {Hernquist}, {Torrey}, {Vogelsberger}, {Weinberger}, \&
  {Marinacci}}]{2018MNRAS.473.4077P}
{Pillepich}, A., {Springel}, V., {Nelson}, D., {et~al.} 2018, \mnras, 473, 4077

\bibitem[{{Piner} {et~al.}(1995){Piner}, {Stone}, \&
  {Teuben}}]{1995ApJ...449..508P}
{Piner}, B.~G., {Stone}, J.~M., \& {Teuben}, P.~J. 1995, \apj, 449, 508

\bibitem[{{Pinna} {et~al.}(2019){Pinna}, {Falc{\'o}n-Barroso}, {Martig},
  {Sarzi}, {Coccato}, {Iodice}, {Corsini}, {de Zeeuw}, {Gadotti}, {Leaman},
  {Lyubenova}, {McDermid}, {Minchev}, {Morelli}, {van de Ven}, \&
  {Viaene}}]{2019A&A...623A..19P}
{Pinna}, F., {Falc{\'o}n-Barroso}, J., {Martig}, M., {et~al.} 2019, \aap, 623,
  A19

\bibitem[{{Pogge}(1989)}]{1989ApJS...71..433P}
{Pogge}, R.~W. 1989, \apjs, 71, 433

\bibitem[{{Prieto} {et~al.}(2019){Prieto}, {Fernandez-Ontiveros}, {Bruzual},
  {Burkert}, {Schartmann}, \& {Charlot}}]{2019MNRAS.485.3264P}
{Prieto}, M.~A., {Fernandez-Ontiveros}, J.~A., {Bruzual}, G., {et~al.} 2019,
  \mnras, 485, 3264

\bibitem[{{Querejeta} {et~al.}(2015){Querejeta}, {Meidt}, {Schinnerer},
  {Cisternas}, {Mu{\~n}oz-Mateos}, {Sheth}, {Knapen}, {van de Ven}, {Norris},
  {Peletier}, {Laurikainen}, {Salo}, {Holwerda}, {Athanassoula}, {Bosma},
  {Groves}, {Ho}, {Gadotti}, {Zaritsky}, {Regan}, {Hinz}, {Gil de Paz},
  {Menendez-Delmestre}, {Seibert}, {Mizusawa}, {Kim}, {Erroz-Ferrer}, {Laine},
  \& {Comer{\'o}n}}]{2015ApJS..219....5Q}
{Querejeta}, M., {Meidt}, S.~E., {Schinnerer}, E., {et~al.} 2015, \apjs, 219, 5

\bibitem[{{Rautiainen} \& {Salo}(2000)}]{2000A&A...362..465R}
{Rautiainen}, P. \& {Salo}, H. 2000, \aap, 362, 465

\bibitem[{{Regan} \& {Teuben}(2004)}]{2004ApJ...600..595R}
{Regan}, M.~W. \& {Teuben}, P.~J. 2004, \apj, 600, 595

\bibitem[{{Roth} {et~al.}(2005){Roth}, {Kelz}, {Fechner}, {Hahn}, {Bauer},
  {Becker}, {B{\"o}hm}, {Christensen}, {Dionies}, {Paschke}, {Popow}, {Wolter},
  {Schmoll}, {Laux}, \& {Altmann}}]{2005PASP..117..620R}
{Roth}, M.~M., {Kelz}, A., {Fechner}, T., {et~al.} 2005, \pasp, 117, 620

\bibitem[{{Sakamoto} {et~al.}(1999){Sakamoto}, {Okumura}, {Ishizuki}, \&
  {Scoville}}]{1999ApJ...525..691S}
{Sakamoto}, K., {Okumura}, S.~K., {Ishizuki}, S., \& {Scoville}, N.~Z. 1999,
  \apj, 525, 691

\bibitem[{{S{\'a}nchez} {et~al.}(2016){S{\'a}nchez}, {Garc{\'\i}a-Benito},
  {Zibetti}, {Walcher}, {Husemann}, {Mendoza}, {Galbany}, {Falc{\'o}n-Barroso},
  {Mast}, {Aceituno}, {Aguerri}, {Alves}, {Amorim}, {Ascasibar},
  {Barrado-Navascues}, {Barrera-Ballesteros}, {Bekerait{\`e}}, {Bland
  -Hawthorn}, {Cano D{\'\i}az}, {Cid Fernandes}, {Cavichia}, {Cortijo},
  {Dannerbauer}, {Demleitner}, {D{\'\i}az}, {Dettmar}, {de
  Lorenzo-C{\'a}ceres}, {del Olmo}, {Galazzi}, {Garc{\'\i}a-Lorenzo}, {Gil de
  Paz}, {Gonz{\'a}lez Delgado}, {Holmes}, {Igl{\'e}sias-P{\'a}ramo}, {Kehrig},
  {Kelz}, {Kennicutt}, {Kleemann}, {Lacerda}, {L{\'o}pez Fern{\'a}ndez},
  {L{\'o}pez S{\'a}nchez}, {Lyubenova}, {Marino}, {M{\'a}rquez},
  {Mendez-Abreu}, {Moll{\'a}}, {Monreal-Ibero}, {Ortega Minakata},
  {Torres-Papaqui}, {P{\'e}rez}, {Rosales-Ortega}, {Roth},
  {S{\'a}nchez-Bl{\'a}zquez}, {Schilling}, {Spekkens}, {Vale Asari}, {van den
  Bosch}, {van de Ven}, {Vilchez}, {Wild}, {Wisotzki}, {Y{\i}ld{\i}r{\i}m}, \&
  {Ziegler}}]{2016A&A...594A..36S}
{S{\'a}nchez}, S.~F., {Garc{\'\i}a-Benito}, R., {Zibetti}, S., {et~al.} 2016,
  \aap, 594, A36

\bibitem[{{Scarlata} {et~al.}(2004){Scarlata}, {Stiavelli}, {Hughes}, {Axon},
  {Alonso-Herrero}, {Atkinson}, {Batcheldor}, {Binney}, {Capetti}, {Carollo},
  {Dressel}, {Gerssen}, {Macchetto}, {Maciejewski}, {Marconi}, {Merrifield},
  {Ruiz}, {Sparks}, {Tsvetanov}, \& {van der Marel}}]{2004AJ....128.1124S}
{Scarlata}, C., {Stiavelli}, M., {Hughes}, M.~A., {et~al.} 2004, \aj, 128, 1124

\bibitem[{{Schaye} {et~al.}(2015){Schaye}, {Crain}, {Bower}, {Furlong},
  {Schaller}, {Theuns}, {Dalla Vecchia}, {Frenk}, {McCarthy}, {Helly},
  {Jenkins}, {Rosas-Guevara}, {White}, {Baes}, {Booth}, {Camps}, {Navarro},
  {Qu}, {Rahmati}, {Sawala}, {Thomas}, \& {Trayford}}]{2015MNRAS.446..521S}
{Schaye}, J., {Crain}, R.~A., {Bower}, R.~G., {et~al.} 2015, \mnras, 446, 521

\bibitem[{{Sellwood} \& {Binney}(2002)}]{2002MNRAS.336..785S}
{Sellwood}, J.~A. \& {Binney}, J.~J. 2002, \mnras, 336, 785

\bibitem[{{Sheth} {et~al.}(2012){Sheth}, {Melbourne}, {Elmegreen}, {Elmegreen},
  {Athanassoula}, {Abraham}, \& {Weiner}}]{2012ApJ...758..136S}
{Sheth}, K., {Melbourne}, J., {Elmegreen}, D.~M., {et~al.} 2012, \apj, 758, 136

\bibitem[{{Sheth} {et~al.}(2010){Sheth}, {Regan}, {Hinz}, {Gil de Paz},
  {Men{\'e}ndez-Delmestre}, {Mu{\~n}oz-Mateos}, {Seibert}, {Kim},
  {Laurikainen}, {Salo}, {Gadotti}, {Laine}, {Mizusawa}, {Armus},
  {Athanassoula}, {Bosma}, {Buta}, {Capak}, {Jarrett}, {Elmegreen},
  {Elmegreen}, {Knapen}, {Koda}, {Helou}, {Ho}, {Madore}, {Masters},
  {Mobasher}, {Ogle}, {Peng}, {Schinnerer}, {Surace}, {Zaritsky},
  {Comer{\'o}n}, {de Swardt}, {Meidt}, {Kasliwal}, \&
  {Aravena}}]{2010PASP..122.1397S}
{Sheth}, K., {Regan}, M., {Hinz}, J.~L., {et~al.} 2010, \pasp, 122, 1397

\bibitem[{{Sheth} {et~al.}(2005){Sheth}, {Vogel}, {Regan}, {Thornley}, \&
  {Teuben}}]{2005ApJ...632..217S}
{Sheth}, K., {Vogel}, S.~N., {Regan}, M.~W., {Thornley}, M.~D., \& {Teuben},
  P.~J. 2005, \apj, 632, 217

\bibitem[{{Shlosman} {et~al.}(1989){Shlosman}, {Frank}, \&
  {Begelman}}]{1989Natur.338...45S}
{Shlosman}, I., {Frank}, J., \& {Begelman}, M.~C. 1989, \nat, 338, 45

\bibitem[{{Sormani} {et~al.}(2015){Sormani}, {Binney}, \&
  {Magorrian}}]{2015MNRAS.449.2421S}
{Sormani}, M.~C., {Binney}, J., \& {Magorrian}, J. 2015, \mnras, 449, 2421

\bibitem[{{Tabor} {et~al.}(2017){Tabor}, {Merrifield}, {Arag{\'o}n-Salamanca},
  {Cappellari}, {Bamford}, \& {Johnston}}]{2017MNRAS.466.2024T}
{Tabor}, M., {Merrifield}, M., {Arag{\'o}n-Salamanca}, A., {et~al.} 2017,
  \mnras, 466, 2024

\bibitem[{{Tabor} {et~al.}(2019){Tabor}, {Merrifield}, {Arag{\'o}n-Salamanca},
  {Fraser-McKelvie}, {Peterken}, {Smethurst}, {Drory}, \&
  {Lane}}]{2019MNRAS.485.1546T}
{Tabor}, M., {Merrifield}, M., {Arag{\'o}n-Salamanca}, A., {et~al.} 2019,
  \mnras, 485, 1546

\bibitem[{{Vazdekis}(1999)}]{1999ApJ...513..224V}
{Vazdekis}, A. 1999, \apj, 513, 224

\bibitem[{{Vazdekis} {et~al.}(2003){Vazdekis}, {Cenarro}, {Gorgas}, {Cardiel},
  \& {Peletier}}]{2003MNRAS.340.1317V}
{Vazdekis}, A., {Cenarro}, A.~J., {Gorgas}, J., {Cardiel}, N., \& {Peletier},
  R.~F. 2003, \mnras, 340, 1317

\bibitem[{{Vazdekis} {et~al.}(2015){Vazdekis}, {Coelho}, {Cassisi},
  {Ricciardelli}, {Falc{\'o}n-Barroso}, {S{\'a}nchez-Bl{\'a}zquez}, {La
  Barbera}, {Beasley}, \& {Pietrinferni}}]{2015MNRAS.449.1177V}
{Vazdekis}, A., {Coelho}, P., {Cassisi}, S., {et~al.} 2015, \mnras, 449, 1177

\bibitem[{{Vazdekis} {et~al.}(2016){Vazdekis}, {Koleva}, {Ricciardelli},
  {R{\"o}ck}, \& {Falc{\'o}n-Barroso}}]{2016MNRAS.463.3409V}
{Vazdekis}, A., {Koleva}, M., {Ricciardelli}, E., {R{\"o}ck}, B., \&
  {Falc{\'o}n-Barroso}, J. 2016, \mnras, 463, 3409

\bibitem[{{Vazdekis} {et~al.}(2010){Vazdekis}, {S{\'a}nchez-Bl{\'a}zquez},
  {Falc{\'o}n-Barroso}, {Cenarro}, {Beasley}, {Cardiel}, {Gorgas}, \&
  {Peletier}}]{2010MNRAS.404.1639V}
{Vazdekis}, A., {S{\'a}nchez-Bl{\'a}zquez}, P., {Falc{\'o}n-Barroso}, J.,
  {et~al.} 2010, \mnras, 404, 1639

\bibitem[{{Verheijen} {et~al.}(2004){Verheijen}, {Bershady}, {Andersen},
  {Swaters}, {Westfall}, {Kelz}, \& {Roth}}]{2004AN....325..151V}
{Verheijen}, M.~A.~W., {Bershady}, M.~A., {Andersen}, D.~R., {et~al.} 2004,
  Astronomische Nachrichten, 325, 151

\bibitem[{{Walker} {et~al.}(1996){Walker}, {Mihos}, \&
  {Hernquist}}]{1996ApJ...460..121W}
{Walker}, I.~R., {Mihos}, J.~C., \& {Hernquist}, L. 1996, \apj, 460, 121

\end{thebibliography}

\begin{appendix} 

\section{Parameter values in each region}
 In this section we present the ranges, mean values and dispersions of the kinematics and stellar population parameters extracted during our analysis in the three regions studied in the galaxies of our sample. The typical uncertainty for \deltaV\, is $7$\,\kms, for \deltaS\, $9$\,\kms, between $0.5$\,Gyr (for the young stellar populations) and $3$\,Gyr \citepads[for the old stellar populations;][]{2019A&A...623A..19P} for mean age, for the fraction of young stars $0.13$, for the colour excess $0.05$, and for [M/H] $0.10$\,dex \citepads[][]{2019A&A...623A..19P}.
 \begin{table*}[!ht]
    \caption{Ranges, means, and dispersions of the kinematics and stellar population parameters extracted in our analysis in the three regions studied in our galaxies. From top to bottom: difference in line-of-sight velocity \deltaV\ , difference in velocity dispersion \deltaS\ , luminosity-weighted mean age, fraction of young stars (YSF), colour excess E(B$-$V), and metallicity [M/H].}
\centering
\begin{tabular}{l|ccc|ccc|ccc}
\hline
\multicolumn{1}{c|}{Region} & \multicolumn{3}{|c|}{Disc} & \multicolumn{3}{|c|}{Ring} & \multicolumn{3}{|c}{Nucleus}\\
\hline
$\Delta v_\star$ ($km\,s^{-1}$) & Range & Mean & Sigma & Range & Mean & Sigma & Range & Mean & Sigma \\
NGC\,1097 & $[0, 25]$ & $5$ & $5$ & $[0, 70]$ & $10$ & $9$ & $[0, 25]$ & $5$ & $4$ \\
NGC\,1300 & $[0, 88]$ & $5$ & $6$ & $[0, 13]$ & $4$ & $3$ & $[0, 12]$ & $3$ & $2$ \\
NGC\,4643 & $[0, 18]$ & $3$ & $3$ & - & - & - & - & - & - \\
NGC\,5248 & $[0, 18]$ & $4$ & $3$ & $[0, 38]$ & $5$ & $5$ & $[0, 14]$ & $4$ & $3$ \\
\hline
$\Delta \sigma_\star$ ($km\,s^{-1}$) & Range & Mean & Sigma & Range & Mean & Sigma & Range & Mean & Sigma \\
NGC\,1097 & $[-11, 53]$ & $10$ & $10$ & $[-38, 88]$ & $24$ & $16$ & $[-14, 37]$ & $5$ & $7$ \\
NGC\,1300 & $[-16, 44]$ & $1$ & $7$ & $[-6, 21]$ & $8$ & $5$ & $[-7, 17]$ & $4$ & $5$ \\
NGC\,4643 & $[-13, 35]$ & $8$ & $6$ & - & - & - & - & - & - \\
NGC\,5248 & $[-10, 35]$ & $6$ & $7$ & $[-22, 46]$ & $11$ & $7$ & $[-1, 23]$ & $9$ & $5$ \\
\hline
Age (Gyr) & Range & Mean & Sigma & Range & Mean & Sigma & Range & Mean & Sigma \\
NGC\,1097 & $[1.4, 11.9]$ & $5.6$ & $1.5$ & $[0.8, 10.7]$ & $3.5$ & $1.8$ & $[2.2, 11.9]$ & $8.1$ & $1.9$ \\
NGC\,1300 & $[4.0, 12.2]$ & $8.3$ & $1.6$ & $[2.3, 6.6]$ & $4.6$ & $0.8$ & $[4.0, 7.4]$ & $5.9$ & $0.7$ \\
NGC\,4643 & $[8.4, 12.6]$ & $11.0$ & $0.7$ & - & - & - & - & - & - \\
NGC\,5248 & $[1.0, 5.0]$ & $3.0$ & $0.7$ & $[0.5, 5.0]$ & $1.8$ & $0.7$ & $[1.5, 4.2]$ & $2.6$ & $0.6$ \\
\hline
YSF & Range & Mean & Sigma & Range & Mean & Sigma & Range & Mean & Sigma \\
NGC\,1097 & $[0.00, 0.79]$ & $0.33$ & $0.18$ & $[0.00, 0.89]$ & $0.44$ & $0.22$ & $[0.00, 0.56]$ & $0.10$ & $0.10$ \\
NGC\,1300 & $[0.00, 0.51]$ & $0.093$ & $0.11$ & $[0.00, 0.50]$ & $0.23$ & $0.11$ & $[0.00, 0.39]$ & $0.19$ & $0.08$ \\
NGC\,4643 & $[0.00, 0.13]$ & $0.00$ & $0.02$ & - & - & - & - & - & - \\
NGC\,5248 & $[0.23, 0.95]$ & $0.53$ & $0.15$ & $[0.27, 0.96]$ & $0.63$ & $0.15$ & $[0.28, 0.86]$ & $0.55$ & $0.16$ \\
\hline
E(B$-$V) & Range & Mean & Sigma & Range & Mean & Sigma & Range & Mean & Sigma \\
NGC\,1097 & $[0.00,0.26]$ & $0.08$ & $0.05$ & $[0.00,0.49]$ & $0.16$ & $0.07$ & $[0.02,0.30]$ & $0.14$ & $0.06$ \\
NGC\,1300 & $[0.00,0.34]$ & $0.14$ & $0.06$ & $[0.11,0.34]$ & 0$.21$ & $0.04$ & $[0.10,0.30]$ & $0.20$ & $0.04$ \\
NGC\,4643 & $[0.00,0.14]$ & $0.00$ & $0.02$ & - & - & - & - & - & - \\
NGC\,5248 & $[0.12,0.33]$ & $0.23$ & $0.04$ & $[0.00,0.35]$ & $0.20$ & $0.05$ & $[0.13,0.30]$ & $0.22$ & $0.04$ \\
\hline
[M/H] (dex) & Range & Mean & Sigma & Range & Mean & Sigma & Range & Mean & Sigma \\
NGC\,1097 & $[-0.40, 0.23]$ & $0.04$ & $0.09$ & $[-1.70, 0.24]$ & $-0.33$ & $0.42$ & $[-0.64, 0.25]$ & $0.15$ & $0.11$ \\
NGC\,1300 & $[-0.62, 0.05]$ & $-0.22$ & $0.08$ & $[-0.44, 0.11]$ & $-0.18$ & $0.12$ & $[-0.26, 0.19]$ & $0.00$ & $0.09$ \\
NGC\,4643 & $[-0.25, 0.27]$ & $-0.24$5 & $0.09$ & - & - & - & - & - & - \\
NGC\,5248 & $[-1.10, 0.25]$ & $-0.09$ & $0.15$ & $[-1.80, 0.24]$ & $-0.21$ & $0.30$ & $[-0.15, 0.30]$ & $0.15$ &$ 0.12$ \\

\label{tab:results}
\end{tabular}
\end{table*}
\newpage

\section{Kinematic differences against stellar population parameters}\label{append: kin_vs_prop}

In the present section we show the possible trends of the kinematic differences between the two selected spectral regions against the stellar population parameters extracted during our analysis. In these plots we separate our results in each region of the galaxies in equally spaced bins for each stellar parameter and estimate the mean value of both the difference in velocity \deltaV\, and the difference in velocity dispersion \deltaS\, in each one of them. The significance of each mean value is analysed by comparing the $1\,\sigma$ confidence interval (represented as a colour-shaded area) with the typical error in our kinematic estimations, $7$\,\kms for \deltaV\, and $9$\,\kms for \deltaS\, (represented as a grey-shaded area).

\begin{figure*}[!ht]
\centering
\includegraphics[width=\textwidth]{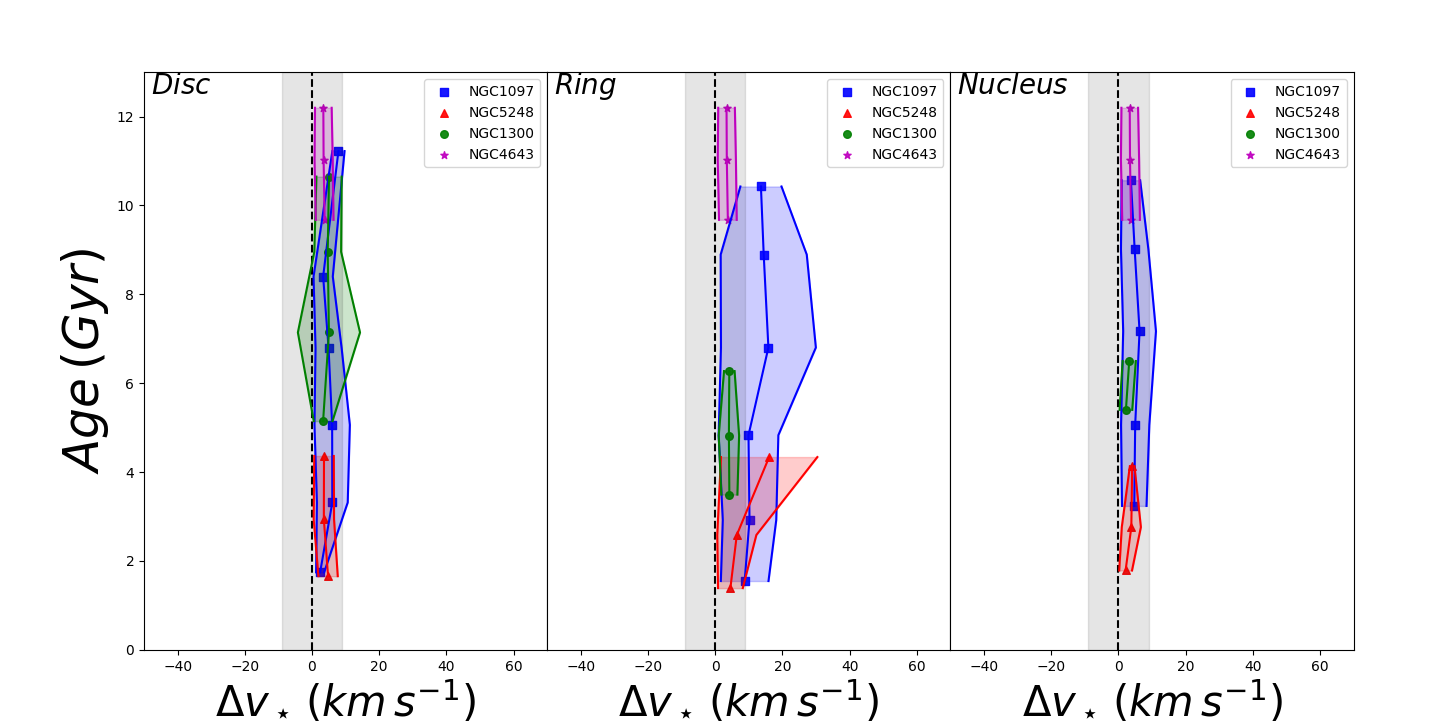}
\caption{Observed values of the difference in blue-red velocities, $\Delta v_\star$, against age in the three selected regions for the four galaxies in the sample. We do not observe any trends of $\Delta v_\star$ with age, except in the case of the ring in NGC\,1097, where the velocity traced by the Ca\,{\sc II} triplet seems to increase with increasing mean ages. Nevertheless, the mean $\Delta v_\star$ value in each bin is in most of the cases not bigger than the typical error. Thus, we conclude that this correlation is not significant. In the case of the other three galaxies we still not observing any trend.}
\label{fig:scat_dv-age}
\end{figure*}

\begin{figure*}[!ht]
\centering
\includegraphics[width=\textwidth]{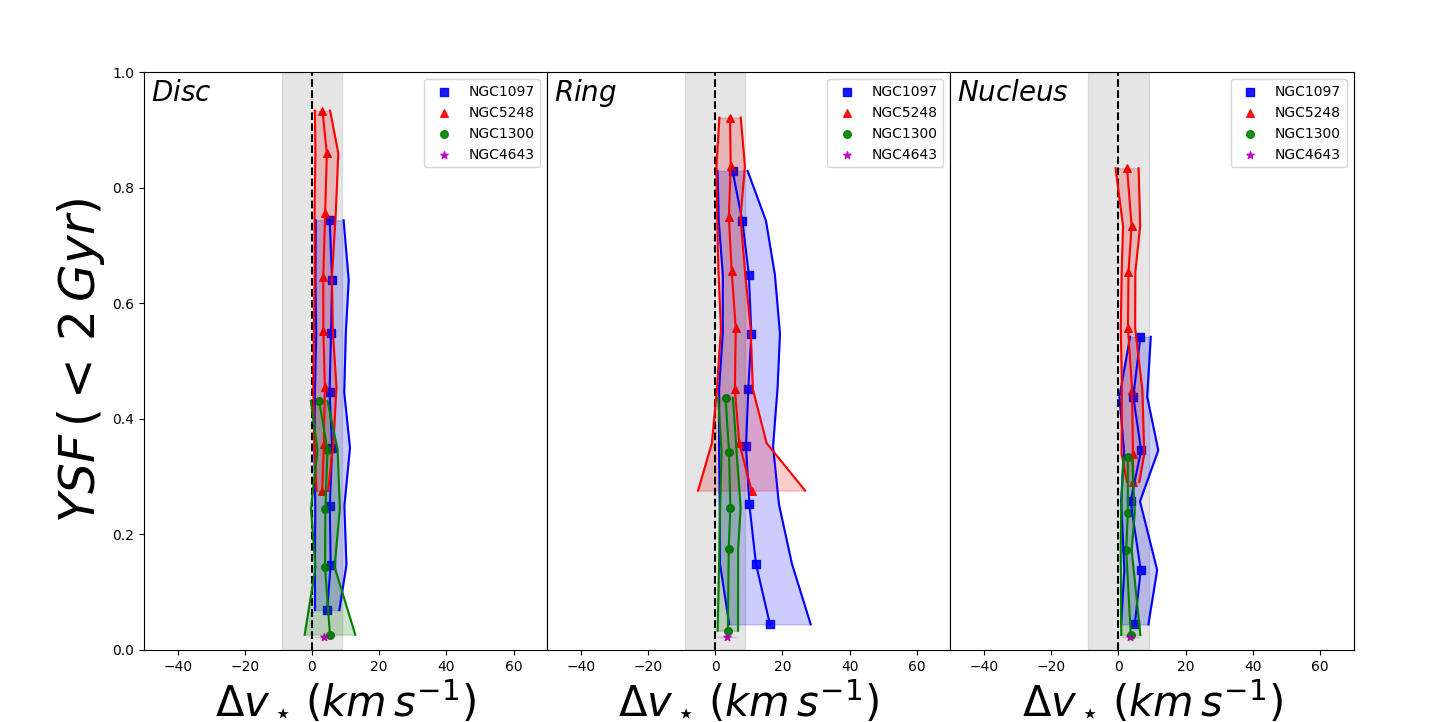}
\caption{As Fig.~\ref{fig:scat_dv-age}, now for \deltaV\, against the fraction of young stars. The scenario is similar to that for the age: we do not have any trends in the discs and nuclei of the four galaxies, which present almost vertical distributions. In the case of the rings we find increasing values of the difference in velocity with increasing values of the fraction of young stars for NGC\,1097, but similarly to what happened with age, the trend is not reliable since the mean values of $\Delta v_\star$ are comparable to the typical error. For the other rings we do not observe any clear trend.}
\label{fig:scat_dv-ysf}
\end{figure*}

\begin{figure*}[!ht]
\centering
\includegraphics[width=\textwidth]{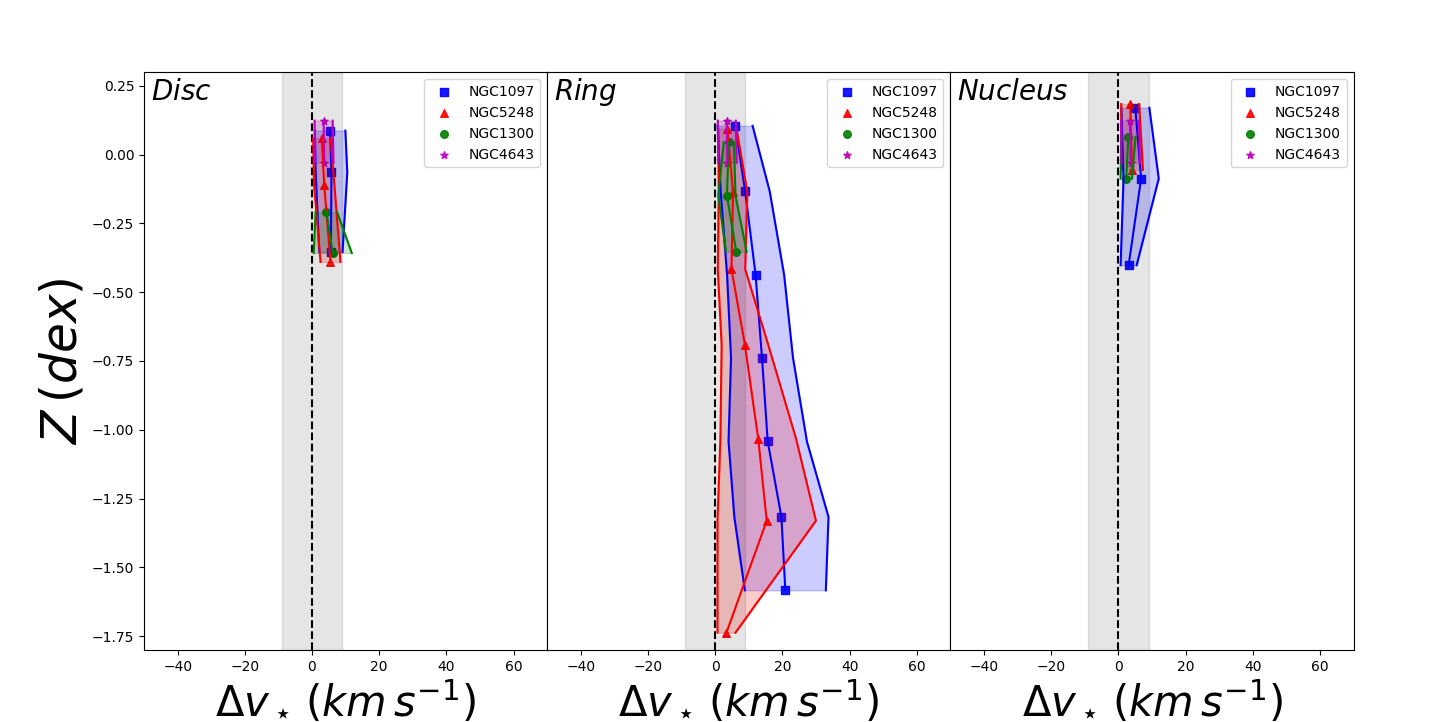}
\caption{As Fig.~\ref{fig:scat_dv-age}, now for \deltaV\, against metallicity [M/H]. The discs and nuclei of the four galaxies span a very small range which complicates the observation of the possible trends. Regarding the rings, only NGC\,1097 and NGC\,5248 span a wide range of values of the metallicity. However, they do not present a clear trend and in most of the bins the mean value of $\Delta v_\star$ is not bigger than the typical error.}
\label{fig:scat_dv-metal}
\end{figure*}

\begin{figure*}[!ht]
\centering
\includegraphics[width=\textwidth]{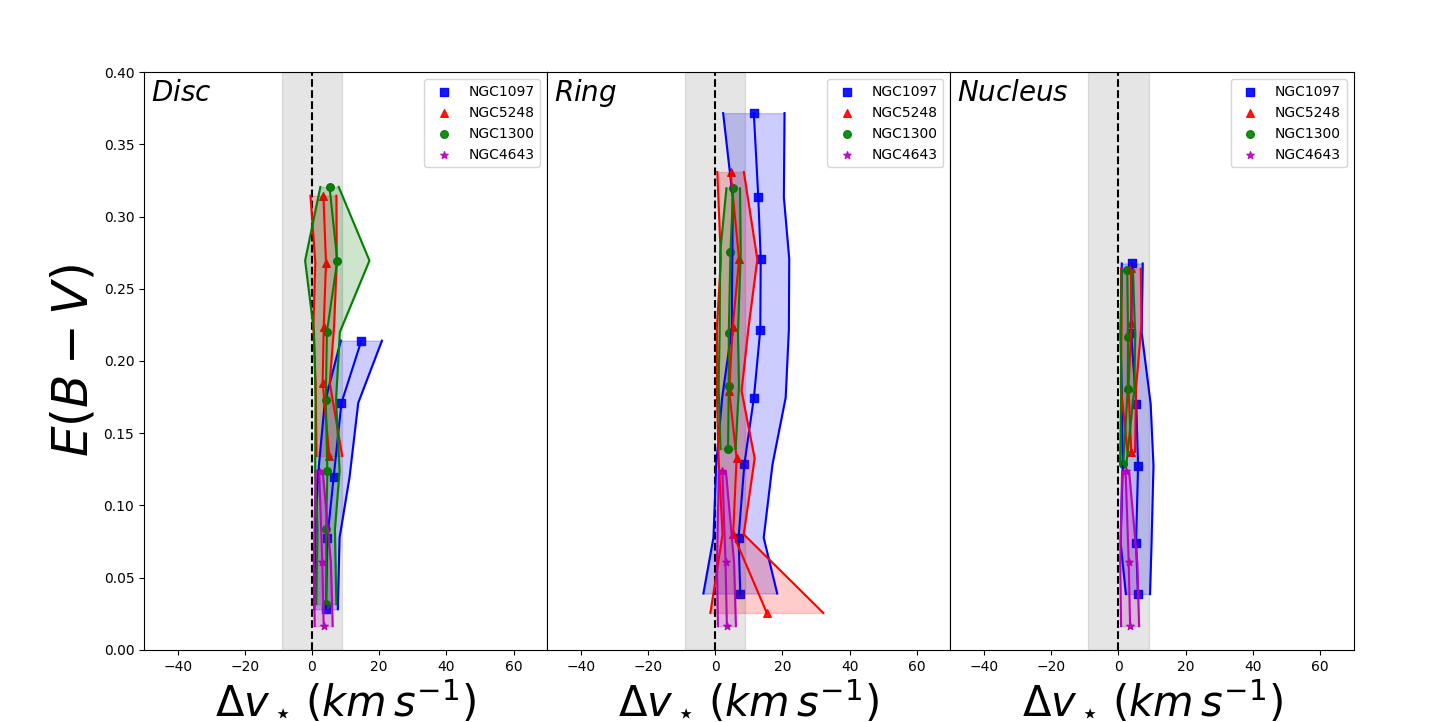}
\caption{As Fig.~\ref{fig:scat_dv-age}, now for \deltaV\, against colour excess E(B$-$V). In general we observe in the three regions of each galaxy almost vertical distributions. The only exception is the disc in NGC\,1097 which present increasing values of $\Delta v_\star$ with increasing values of E(B$-$V). Nevertheless, as we previously stated for other parameters, the trend is not reliable since the mean values of $\Delta v_\star$ in each bin are comparable to the value of the typical error.}
\label{fig:scat_dv-ebv}
\end{figure*}

\begin{figure*}[!ht]
\centering
\includegraphics[width=\textwidth]{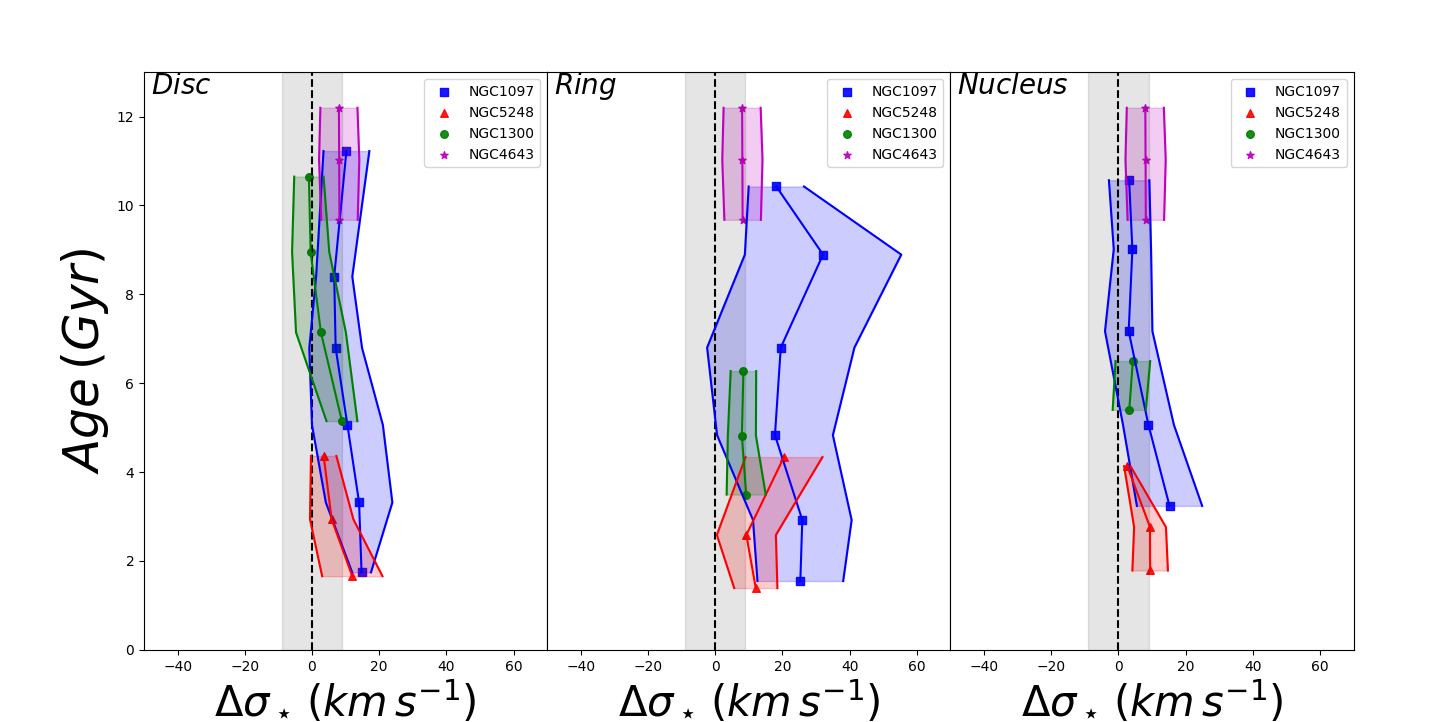}
\caption{As Fig.~\ref{fig:scat_dv-age}, now for \deltaS\, against age. In general we argue that the three regions of each galaxy span little age range, which hinder the study of possible trends. The only exceptions are the three regions of NGC\,1097 and the disc of NGC\,1300. In the cases of the disc in NGC\,1300 and the disc and nucleus in NGC\,1097 we find vertical distributions, indicating the absence of any trend. For the ring in NGC\,1097 we observe high mean values of $\Delta\sigma_\star$ when compared with the typical error, but we do not find a clear distribution pattern, in addition to the fact that they present wide $1\,\sigma$ uncertainty intervals.}
\label{fig:scat_dsig-age}
\end{figure*}

\begin{figure*}[!ht]
\centering
\includegraphics[width=\textwidth]{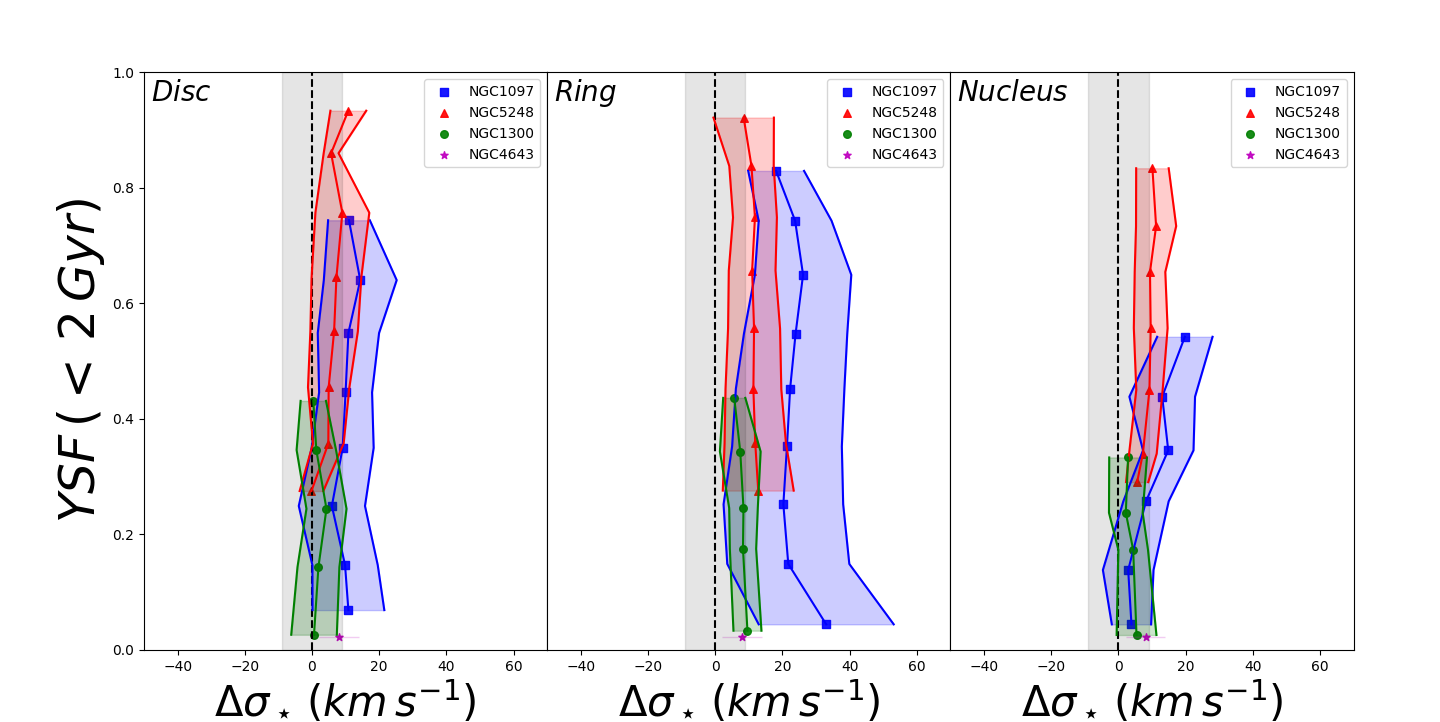}
\caption{As Fig.~\ref{fig:scat_dv-age}, now for \deltaS\, against the fraction of young stars. Similarly to the case of $\Delta v_\star$ against the fraction of young stars we find almost vertical distributions of the mean values of $\Delta\sigma_\star$ in all the regions of each galaxy. The only exception is the nucleus in NGC\,1097 where we find a slight trend of increasing velocity dispersion differences with increasing fractions, but we find mean values of $\Delta\sigma_\star$ very similar to the typical error. Thus, we rule out any kind of trend with the fraction of young stars.}
\label{fig:scat_dsig-ysf}
\end{figure*}

\begin{figure*}[!ht]
\centering
\includegraphics[width=\textwidth]{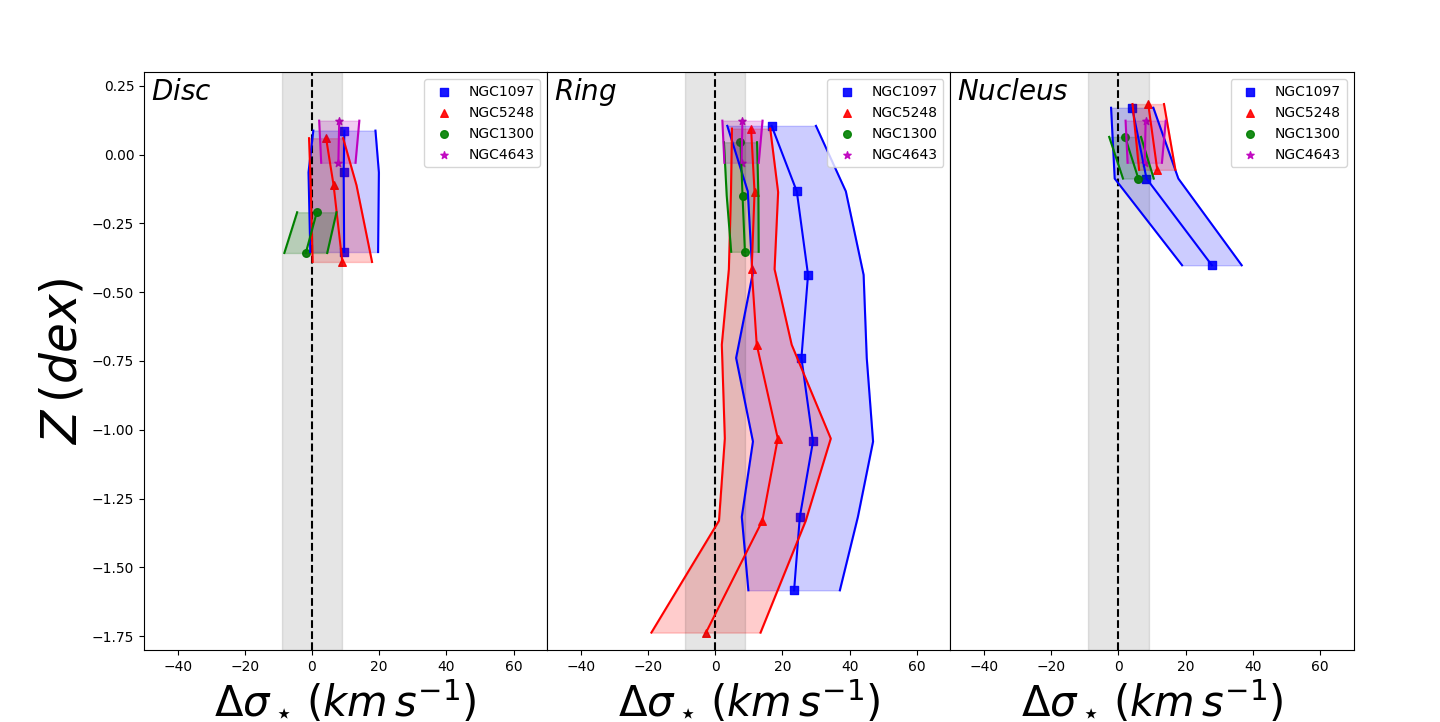}
\caption{As Fig.~\ref{fig:scat_dv-age}, now for \deltaS\, against metallicity [M/H]. In the cases of the discs and nuclei of the whole sample we find not enough bins to discern a clear trend. The rings in NGC\,1097 and NGC\,5248 span a wide range of values of the metallicity, but they present a vertical distribution of the mean values of $\Delta\sigma_\star$, indicating the absence of a trend.}
\label{fig:scat_dsig-metal}
\end{figure*}

\section{Kinematic differences in NGC\,1300 and NGC\,5248}\label{append: kin_1300_5248}

Although NGC\,1300 and NGC\,5248 do not show clear trends of the differences in kinematics with the stellar population parameters, it is still interesting to look for possible trends. In this sense the representation of the whole set of data in each regions can provide us with valuable information.

\begin{figure}[!ht]
\centering
\includegraphics[scale=0.5]{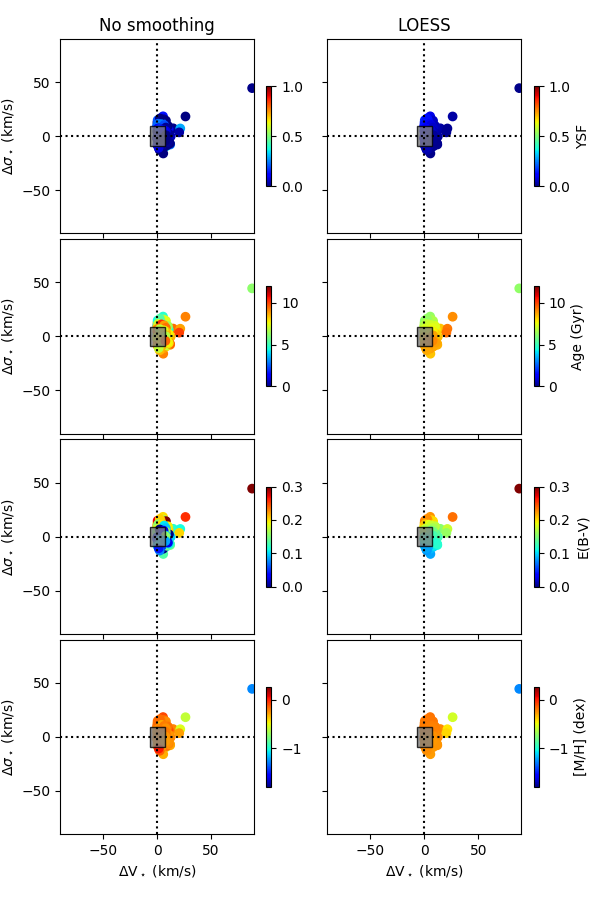}
\caption{$\Delta\sigma_\star$ against $\Delta v_\star$ with colour-coded values of the stellar population parameters (from top to bottom: fraction of young stars, mean age, colour excess, and metallicity) for the disc in NGC\,1300. On the left side we represent the data with no smoothing and on the right side the \textsc{loess}-smoothed data. The grey rectangle represents the typical error in $\Delta v_\star$ ($7$\kms) and $\Delta\sigma_\star$ ($9$\kms). The uncertainties introduced by the smoothing are: $\pm 0.11$ for the YSF, $\pm 1.4$\,Gyr for the mean age, $\pm 0.05$ for E(B$-$V), and $\pm 0.07$\,dex for the metallicity.}
\label{fig:ppxf_out_1300}
\end{figure}

\begin{figure}[!ht]
\centering
\includegraphics[scale=0.5]{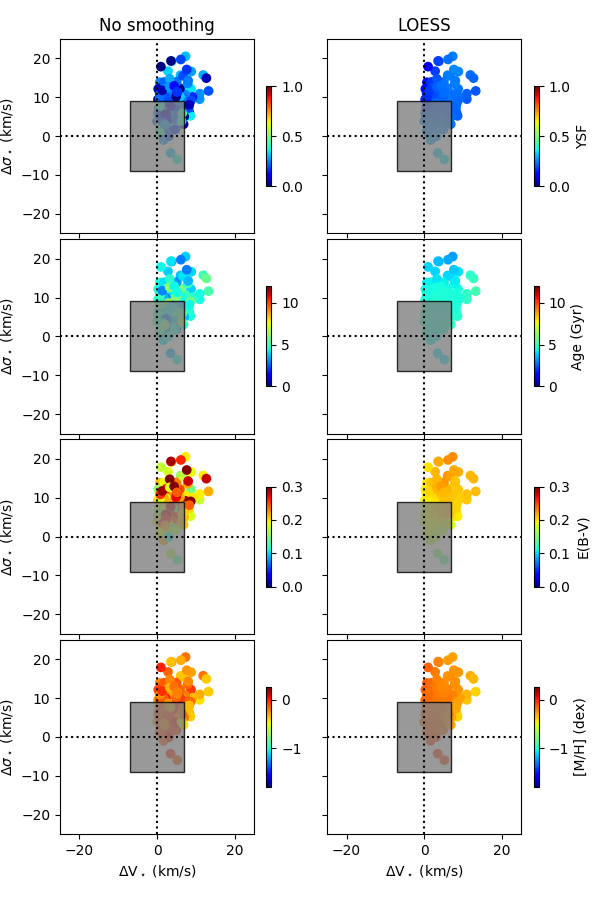}
\caption{As Fig.~\ref{fig:ppxf_out_1300}, now for the ring in NGC\,1300. The uncertainties introduced by the smoothing are: $\pm 0.11$ for the YSF, $\pm 0.7$\,Gyr for the mean age, $\pm 0.04$ for E(B$-$V), and $\pm 0.10$\,dex for the metallicity.}
\label{fig:ppxf_ring_1300}
\end{figure}

\begin{figure}[!ht]
\centering
\includegraphics[scale=0.5]{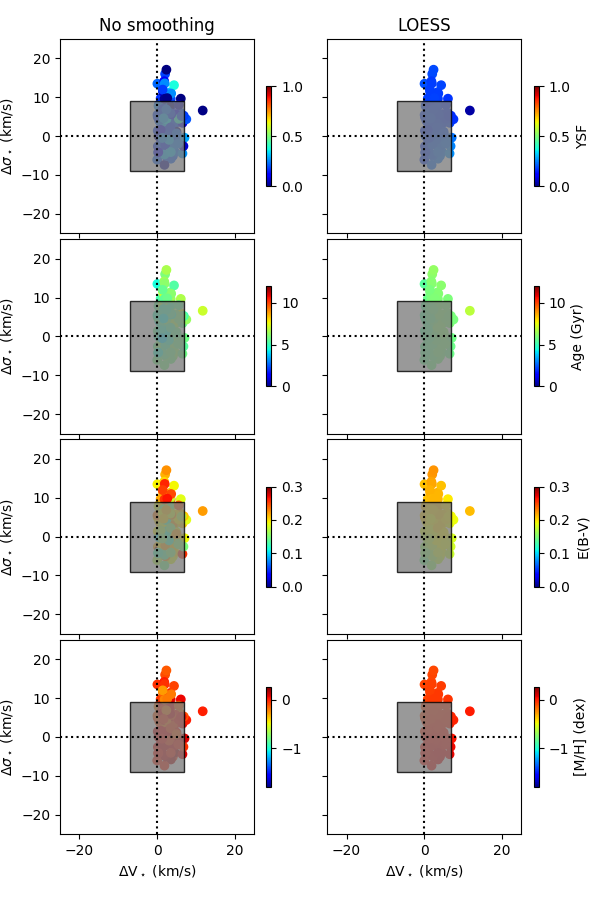}
\caption{As Fig.~\ref{fig:ppxf_out_1300}, now for the nucleus in NGC\,1300. The uncertainties introduced by the smoothing are: $\pm 0.08$ for the YSF, $\pm 0.6$\,Gyr for the mean age, $\pm 0.03$ for E(B$-$V), and $\pm 0.08$\,dex for the metallicity.}
\label{fig:ppxf_nuc_1300}
\end{figure}

All three regions in NGC\,1300 (see Fig.~\ref{fig:ppxf_out_1300} for the disc, Fig.~\ref{fig:ppxf_ring_1300} for the ring, and Fig.~\ref{fig:ppxf_nuc_1300} for the nucleus) exhibit rather low values of $\Delta v_\star$ and $\Delta\sigma_\star$, close to the values of the typical uncertainties. We find that the dynamical ranges are quite small and comparable to the uncertainties in most of the cases. However, we glimpse trends of $\Delta\sigma_\star$ with mean age in the disc and with colour excess in the disc and ring. In all cases the range of values is small, especially in the case of mean age.

\begin{figure}[!ht]
\centering
\includegraphics[scale=0.5]{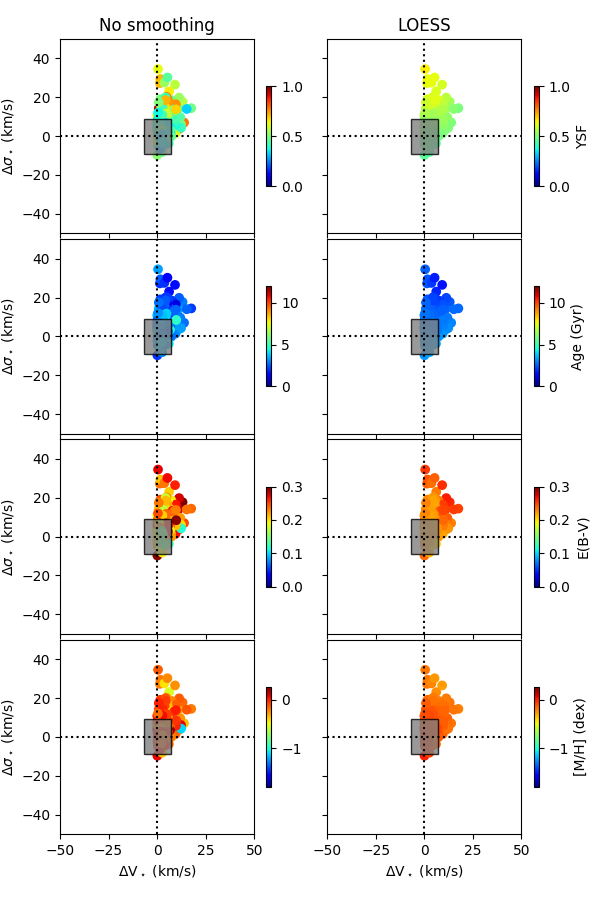}
\caption{As Fig.~\ref{fig:ppxf_out_1300}, now for the disc in NGC\,5248. The uncertainties introduced by the smoothing are: $\pm 0.14$ for the YSF, $\pm 0.6$\,Gyr for the mean age, $\pm 0.03$ for E(B$-$V), and $\pm 0.14$\,dex for the metallicity.}
\label{fig:ppxf_out_5248}
\end{figure}

\begin{figure}[!ht]
\centering
\includegraphics[scale=0.5]{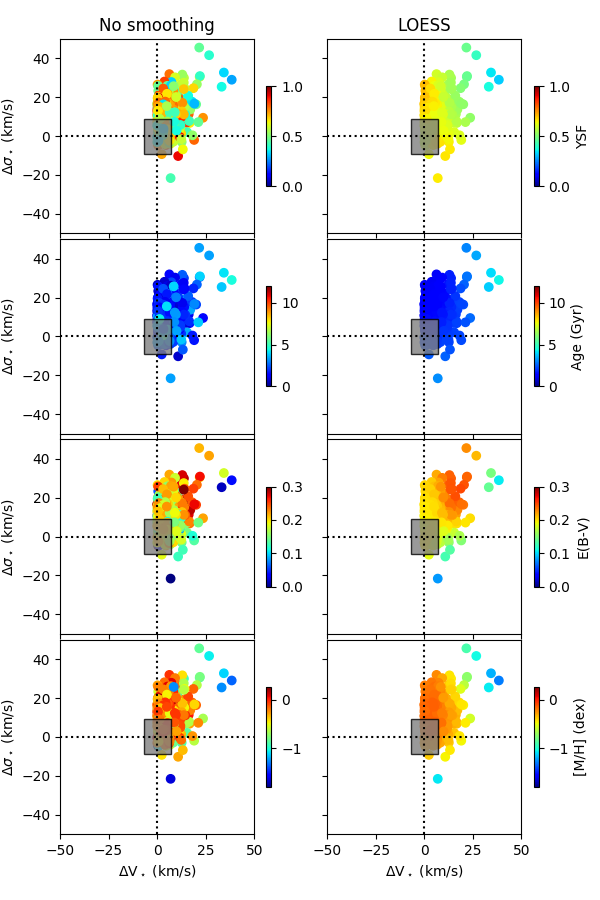}
\caption{As Fig.~\ref{fig:ppxf_out_1300}, now for the ring in NGC\,5248. The uncertainties introduced by the smoothing are: $\pm 0.14$ for the YSF, $\pm 0.7$\,Gyr for the mean age, $\pm 0.05$ for E(B$-$V), and $\pm 0.3$\,dex for the metallicity.}
\label{fig:ppxf_ring_5248}
\end{figure}

\begin{figure}[!ht]
\centering
\includegraphics[scale=0.5]{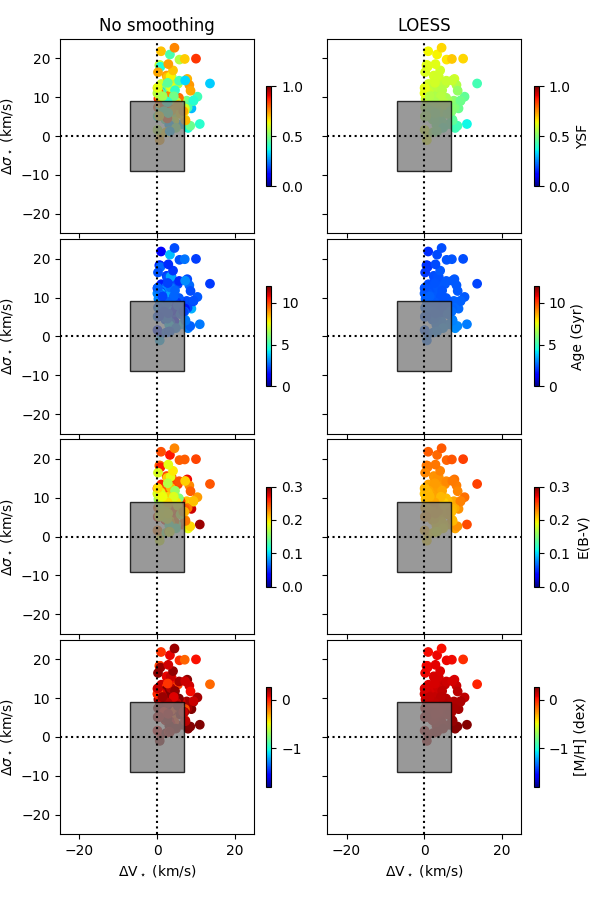}
\caption{As Fig.~\ref{fig:ppxf_out_1300}, now for the nucleus in NGC\,5248. The uncertainties introduced by the smoothing are: $\pm 0.14$ for the YSF, $\pm 0.6$\,Gyr for the mean age, $\pm 0.03$ for E(B$-$V), and $\pm 0.10$\,dex for the metallicity.}
\label{fig:ppxf_nuc_5248}
\end{figure}

The case of NGC\,5248 (see Fig.~\ref{fig:ppxf_out_5248} for the disc, Fig.~\ref{fig:ppxf_ring_5248} for the ring, and Fig.~\ref{fig:ppxf_nuc_5248} for the nucleus) is slightly different. Here $\Delta v_\star$ and $\Delta\sigma_\star$ span wider ranges of values. On the other hand, most of the stellar population parameters show small ranges. In the case of the ring we discern a correlation of $\Delta\sigma_\star$ with colour excess which is similar but not as strong as the one in NGC\,1097.

\begin{figure}[!ht]
\centering
\includegraphics[scale=0.5]{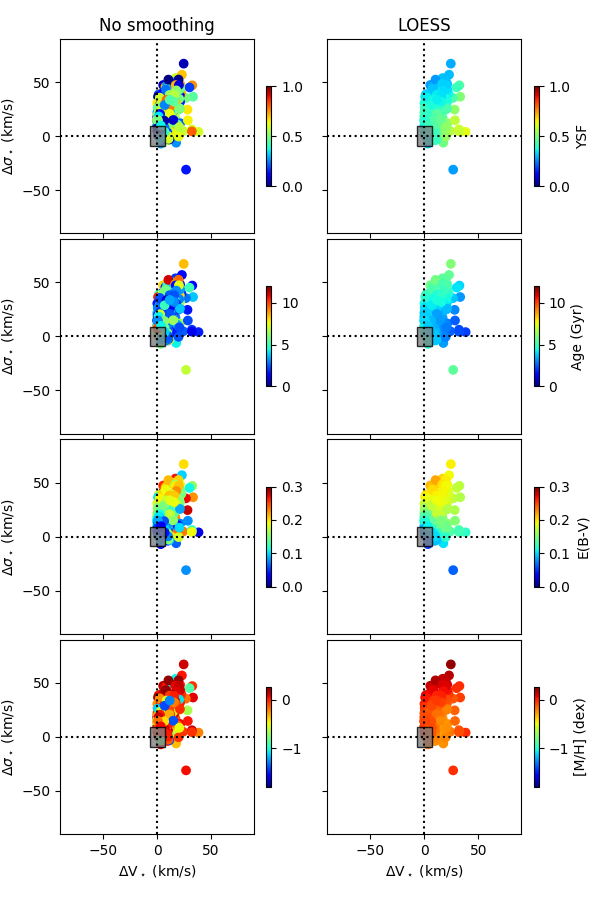}
\caption{As Fig.~\ref{fig:ppxf_out_1300}, now for the ring in NGC\,1097 with a binning level at a target SNR of $250$. The uncertainties introduced by the smoothing are: $\pm 0.23$ for the YSF, $\pm 2.3$\,Gyr for the mean age, $\pm 0.05$ for E(B$-$V), and $\pm 0.35$\,dex for the metallicity.}
\label{fig:ppxf_ring_1097_25}
\end{figure}

\begin{figure}[!ht]
\centering
\includegraphics[scale=0.5]{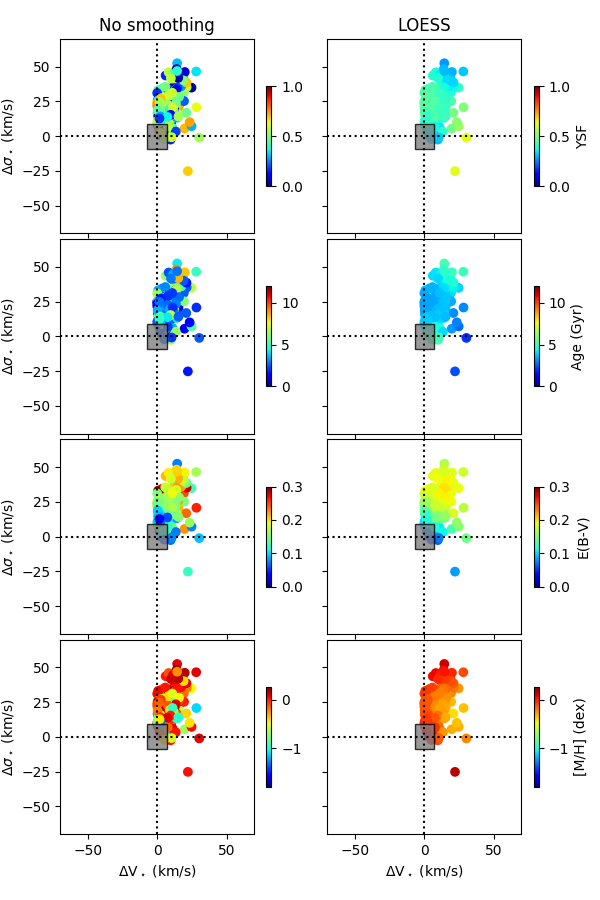}
\caption{As Fig.~\ref{fig:ppxf_out_1300}, now for the ring in NGC\,1097 with a binning level at a target SNR of $350$. The uncertainties introduced by the smoothing are: $\pm 0.21$ for the YSF, $\pm 1.9$\,Gyr for the mean age, $\pm 0.07$ for E(B$-$V), and $\pm 0.3$\,dex for the metallicity.}
\label{fig:ppxf_ring_1097_35}
\end{figure}

In Fig.~\ref{fig:ppxf_ring_1097_25} and Fig.~\ref{fig:ppxf_ring_1097_35} we plot the complete cloud of points corresponding to the analysis of the ring of NGC\,1097 at target signal-to-noise ratios of 250 and 350, respectively. These changes in the SNR are translated into changes in the average size of the bins in the ring. Considering a distance of $15.2{\rm\,Mpc}$ \citepads{2003A&A...412...45P} we have bin sizes of $22{\rm\,pc}$ at SNR=$100$, $63{\rm\,pc}$ at SNR=$250$, and $73{\rm\,pc}$ at SNR=$350$. We see how the kinematic differences decrease as the SNR is increased. Specifically, the ranges in $\Delta v_\star$ and $\Delta\sigma_\star$ are around 50\% and 40\% smaller at SNR=$350$ than at SNR=$100$, respectively.

In the rings of NGC\,1300 and NGC\,5248, at SNR=$100$, we have average bin sizes of $39{\rm\,pc}$ and $35{\rm\,pc}$, at distances of $20.2{\rm\,Mpc}$ and $17.9{\rm\,Mpc}$ \citepads{2003A&A...412...45P}. This means that we will be losing the extreme values in the rings of these galaxies, similarly to what happens when the bin size is increased in NGC\,1097.

The size of the bin thus plays some role in the prevalence of the extreme values of the kinematic differences. The current generation of IFUs is able to detect these values even at reasonable levels of SNR. It is likely that the next generation of IFUs, like VIRUS-W \citepads[][]{2008SPIE.7014E..73F}, MEGARA \citepads[][]{2018SPIE10702E..17G}, or WEAVE-LIFU \citepads[][]{2018SPIE10702E..1BD}, will produce data with enough resolution to properly separate different stellar populations.

\end{appendix}

\end{document}